\documentclass[twocolumn,numberedappendix,twocolappendix]{openjournal}

\usepackage{xcolor}
\usepackage{textgreek}
\usepackage[utf8]{inputenc}
\usepackage{amsmath}
\usepackage{bm}
\usepackage[english]{babel}
\usepackage{xspace}

\usepackage{hyperref}
\hypersetup{
    unicode,
    colorlinks=true,
    linkcolor=linkcolor,
    citecolor=linkcolor,
    filecolor=linkcolor,
    urlcolor=linkcolor,
}
\usepackage{cleveref}
\AtBeginDocument{%
}
\crefname{figure}{Fig.}{Figs.}
\crefname{table}{Table}{Tables}
\makeatletter
\renewcommand{\figurename}{Fig.}
\renewcommand{\tablename}{Table}
\renewcommand{\fnum@figure}{\figurename~\thefigure}
\renewcommand{\fnum@table}{\tablename~\thetable}
\makeatother
\usepackage{color,colortbl}
\definecolor{linkcolor}{rgb}{0.0,0.3,0.5}
\usepackage{tensind}
\tensordelimiter{?}
\DeclareGraphicsExtensions{.bmp,.png,.jpg,.pdf}
\usepackage{verbatim}
\usepackage[normalem]{ulem}
\usepackage{orcidlink}
\usepackage{soul}
\usepackage{acro}
\acsetup{patch/longtable=false}

\DeclareAcronym{BORG}{short = \texttt{BORG}, long  = \textit{Bayesian Origin Reconstruction from Galaxies}}
\DeclareAcronym{CMB}{short = CMB, long  = cosmic microwave background}
\DeclareAcronym{LG}{short = LG, long = Local Group}
\DeclareAcronym{tSZ}{short = tSZ, long  = thermal Sunyaev-Zel'dovich}
\DeclareAcronym{HMC}{short = HMC, long = Hamiltonian Monte Carlo}
\DeclareAcronym{NUTS}{short = NUTS, long = No-U-Turn Sampler}
\DeclareAcronym{CDM}{short = CDM, long = cold dark matter}
\DeclareAcronym{LCDM}{short = $\Lambda$CDM, long  = $\Lambda$-cold dark matter}
\DeclareAcronym{GA}{short = GA, long = \textit{Great Attractor}}
\DeclareAcronym{cGA}{short = cGA, long = Classical \textit{Great Attractor}}
\DeclareAcronym{sGA}{short = sGA, long = Streamline \textit{Great Attractor}}
\DeclareAcronym{dGA}{short = dGA, long = Dynamical \textit{Great Attractor}}
\DeclareAcronym{BPT}{short = \texttt{BPT}, long = \texttt{Beyond-Present-Time}}
\DeclareAcronym{ZoA}{short = ZoA, long = Zone of Avoidance}

\newcommand{\Mpch}{\ensuremath{h^{-1}\,\mathrm{Mpc}}}
\newcommand{\Mpc}{\ensuremath{\mathrm{Mpc}}}
\newcommand{\cMpch}{\ensuremath{h^{-1}\,\mathrm{cMpc}}}

\newcommand{\Msunh}{\ensuremath{h^{-1}\,M_\odot}}
\newcommand{\Msun}{\ensuremath{\mathrm{M}_\odot}}
\newcommand{\kmsec}{\ensuremath{\mathrm{km}\,\mathrm{s}^{-1}}}

\newcommand{\Manti}{\texttt{Manticore-Local}\xspace}
\newcommand{\TWOMPP}{2M\texttt{++}\xspace}

\graphicspath{ {./figs/} }

\begin{document}
\title{Revisiting the Great Attractor: The Local Group's streamline trajectory, cosmic velocity and dynamical fate}

\author{Richard Stiskalek\orcidlink{0000-0002-0986-314X}}
\email{richard.stiskalek@physics.ox.ac.uk}
\affiliation{Astrophysics, University of Oxford, Denys Wilkinson Building, Keble Road, Oxford, OX1 3RH, UK}

\author{Harry Desmond\orcidlink{0000-0003-0685-9791}}
\affiliation{Institute of Cosmology \& Gravitation, University of Portsmouth, Dennis Sciama Building, Portsmouth, PO1 3FX, UK}

\author{Stuart McAlpine\orcidlink{0000-0002-8286-7809}}
\affiliation{The Oskar Klein Centre, Department of Physics, Stockholm University, Albanova University Center, 106 91 Stockholm, Sweden}

\author{Guilhem Lavaux\orcidlink{0000-0003-0143-8891}}
\affiliation{CNRS \& Sorbonne Universit\'e, Institut d'Astrophysique de Paris (IAP), UMR 7095, 98 bis bd Arago, F-75014 Paris, France}

\author{Jens Jasche\orcidlink{0000-0002-4677-5843}}
\affiliation{The Oskar Klein Centre, Department of Physics, Stockholm University, Albanova University Center, 106 91 Stockholm, Sweden}

\author{Michael J. Hudson\orcidlink{0000-0002-1437-3786}}
\affiliation{
Department of Physics and Astronomy, University of Waterloo, Waterloo, ON N2L 3G1, Canada\\
Waterloo Centre for Astrophysics, University of Waterloo, Waterloo, ON N2L 3G1, Canada\\
Perimeter Institute for Theoretical Physics, Waterloo, ON N2L 2Y5, Canada}

\begin{abstract}
We revisit the Great Attractor using the \Manti suite of digital twins of the nearby Universe.
The Great Attractor concept has been proposed as an answer to three distinct questions: what sources the Local Group velocity in the cosmic microwave background frame, where present-day velocity streamlines converge, and where the Local Group is moving to.
Addressing the original motivation of the Great Attractor---explaining the Local Group cosmic velocity---we find that mass within $155~\Mpch$ accounts for only ${\sim}72\%$ of that velocity magnitude with ${\sim}38\,\deg$ directional offset. We show that even in the purely linear regime convergence within this volume is not guaranteed, particularly when also accounting for small-scale contributions to the observer velocity; no single structure, including the proposed Great Attractor, would be expected to dominate the velocity budget.
Streamline convergence is smoothing-scale-dependent, transitioning from Virgo at small scales through the Hydra--Centaurus region at intermediate scales to Shapley at large scales; at intermediate smoothing the convergence point lies near Abell 3565 with an asymmetric basin of mass $\log( M / (\Msunh)) = 16.4 \pm 0.1$ that excludes Norma.
To address the third question, we evolve the \Manti realisations to scale factor $a = 10$ in a new \texttt{Beyond-Present-Time} simulation suite and identify the asymptotic future location of the Local Group.
We find that the dominant motion is towards Virgo, but even it contributes at most one third of the Local Group velocity.
Our results demonstrate that the classical Great Attractor is not a dynamically dominant structure but an artefact of the instantaneous velocity field, and that no single attractor is likely to account for the Local Group motion in the cosmic rest frame.
\end{abstract}


\maketitle

\section{Introduction}\label{sec:intro}

A key component of observational astrophysics is cosmography, the mapping out of nearby structures in the Universe.
While this is most readily done using the luminosities at various wavelengths of galaxies and clusters, complementary information is provided by galaxies' \emph{velocities}, which encode their trajectories and hence provide information about their past and future.
This is necessary to build a detailed picture of the dynamics of the local Universe and the components thereof.

An important role in this endeavour is played by \emph{attractors}, massive objects or collections of objects which dominate the gravitational field in large volumes of space.
As hierarchical structure formation proceeds, such objects become richer over time while surrounding regions---the voids---become emptier.
The most obvious attractors are the superclusters readily visible to optical telescopes, including Virgo, Coma and Shapley.
However, other attractors are more difficult to identify, for example because they are located in the \ac{ZoA} behind the disk of the Milky Way, where dust obscuration makes optical astronomy difficult.
Velocity information provides an effective means of identifying such structures, as infall toward them can be detected in unobscured neighbouring regions~\citep{Lilje_1986,Bell_1988,Pomarede_2017,KraanKorteweg_2017,Courtois_2019,Pomarede_2020}.

An obscured structure of particular interest is the \ac{GA}, the focus of this paper.
The \ac{GA} emerged in the late 1980s as a large-scale density enhancement inferred from peculiar-velocity surveys of galaxies, motivated by the need to explain the peculiar velocity of the \ac{LG} with respect to the \ac{CMB}~\citep{Smoot}.
\citet{Dressler_1987} measured peculiar velocities of elliptical galaxies and fitted a bulk flow, noting that the Hydra--Centaurus region itself appeared to be moving toward mass concentrations beyond ${\sim}5000~\kmsec$.
Earlier work by~\citet{Aaronson_1982} had already identified deviations from pure Hubble expansion, though without locating a definitive source for the observed motion.
Similarly,~\citet{Lilje_1986} detected a significant quadrupolar tidal velocity field in the Local Supercluster caused by external density structure, with the dominant eigenvector pointing toward Hydra--Centaurus.
Building on this,~\citet{Bell_1988} fitted an attractor model and identified the \ac{GA} as a large-scale overdensity at $(\ell,\,b) \approx (307,\,9)\,\deg$ at a distance of $43.5~\Mpch$, located beyond Hydra--Centaurus in order to explain its motion. We shall later refer to this as the ``classical'' Great Attractor.

Subsequent analyses refined this picture.~\citet{Hudson_1993_A} mapped the density field from optical galaxies, showing that the overdensity in the Centaurus--Hydra--Pavo--Virgo supercluster complex peaks near the Centaurus cluster (Cen~30) at a distance of $35~\Mpch$, not at the \ac{GA} distance of $43.5~\Mpch$.~\citet{Hudson_1993_B} used this density field to predict the \ac{LG} peculiar velocity, finding a ${\sim}30\,\deg$ misalignment with the \ac{CMB} dipole direction.
\citet{Hudson_1994_D} concluded that there is no evidence for a \ac{GA} within $80~\Mpch$ being primarily responsible for the motion of nearby galaxies, with $\gtrsim 400~\kmsec$ attributable to sources beyond this distance.
This density-field approach, mapping the galaxy distribution and predicting peculiar velocities via linear theory, was further developed by~\citet{Pike_2005} and~\citet{Carrick_2015}, and extended to full Bayesian inference of the initial conditions by~\citet{Jasche_2019}, providing the foundation for the digital twins used in this work.
However, a large portion of the \ac{GA} region lies behind the \ac{ZoA}, making direct optical observation of galaxies difficult due to dust obscuration and stellar confusion and hence necessitating velocity-based analysis.
In this context, the \ac{GA} came to be understood not as the sole driver of the \ac{CMB} dipole, but as an important intermediate overdensity embedded within a larger-scale flow pattern extending to greater depths.

Despite these challenges, subsequent measurements reinforced the notion of a coherent inflow to the \ac{GA} region.
Using Fundamental Plane and Tully--Fisher distance indicators,~\citet{Dressler_1990, Dressler_1990B} and~\citet{Burstein_1990} found evidence for a convergence of peculiar velocities near $4000$--$4500~\kmsec$, consistent with the inferred distance to the \ac{GA}.
However,~\citet{Hudson_1994_C} showed that inhomogeneous Malmquist bias can mimic such convergence; after correcting for this effect, the data are better described by infall of ${\sim}240~\kmsec$ onto Cen~30 combined with a residual bulk dipole of ${\sim}360~\kmsec$, with no need for convergence at the \ac{GA} distance.
Nevertheless, the amplitude and extent of the measured bulk flows posed a significant challenge to \ac{CDM} models at the time.
$N$-body simulations by~\citet{Bertschinger_1988} showed that the mass concentration required to produce such motions was improbable in \ac{CDM} scenarios assuming standard galaxy biasing, suggesting that the \ac{GA} could impose non-trivial constraints on large-scale structure formation models.

The current understanding is that the \ac{GA} is not a gravitationally bound object but a diffuse overdense region influencing galaxy motions on ${\sim} 100~\Mpch$ scales.
Within this region, the Norma cluster (Abell 3627) has been thought to be a plausible core.
Identified through HI and near-infrared observations obscured by the Milky Way, Norma was found to be a rich and massive cluster located at $(\ell,\,b)\approx(325,\,-7)\,\deg$ and $cz\approx 4870~\kmsec$~\citep{Kraan-Korteweg_1996}.
Subsequent dynamical analysis by~\citet{Woudt_2008} confirmed its significance, though the \ac{GA}'s mass distribution does not converge neatly on Norma, suggesting a broader structure.
Another massive cluster hidden behind the \ac{ZoA} is the Vela supercluster, though it is considerably further away than the \ac{GA} with a recessional velocity of $18{,}000~\kmsec$~\citep{KraanKorteweg_2017,Courtois_2019,Hatamkhani_2023,Rajohnson_2024}.
The South Pole Wall, at $cz \approx 12{,}000~\kmsec$, is another large-scale structure detected through peculiar-velocity analysis~\citep{Pomarede_2020}.

Several studies pointed to the necessity of considering other large-scale overdensities in understanding the peculiar motion of the \ac{LG}.
The Perseus--Pisces supercluster was identified as another prominent structure contributing comparably to the density field within the same volume.
In some interpretations, Perseus--Pisces itself appears to be moving in the direction of the \ac{GA}, implying the influence of more massive structures beyond, such as the Shapley Supercluster~\citep{Scaramella_1989,Raychaudhury_1989}.
These findings challenged the notion of the \ac{GA} as a unique attractor and instead indicated that it may form part of a broader hierarchy of flows, a picture that our results will further substantiate.

Modern reconstructions of the large-scale velocity field, particularly with the CosmicFlows programme, have offered a more nuanced picture~\citep{Courtois_2013,Tully_2014,CF3_cosmography,Courtois_2023,Tully_2023,Hoffman_2024,Courtois_2025}.
They have shifted the definition of the \ac{GA}: whereas the original formulation by~\citet{Bell_1988} sought a massive overdensity whose gravitational influence would explain the \ac{LG}'s peculiar velocity, more recent work defines the \ac{GA} through streamline convergence and the associated \emph{watershed basin} within which velocities flow toward a common attractor. \citet{Courtois_2013} first presented velocity-field streamlines in the \ac{GA} region, and~\citet{Tully_2014} subsequently used such streamlines to define watershed basins.
\citet{Valade_2024} used Hamiltonian Monte Carlo techniques on CosmicFlows-4 data to identify basins of attraction out to $30{,}000~\kmsec$, finding a slight preference for placing the \ac{GA} within the basin of the more massive Shapley attractor. Their analysis identifies the dominant basin with the Sloan Great Wall, occupying a volume of $15.5 \times 10^6~(\!\Mpch)^3$.~\citet{Dupuy_2025} develop a deep-learning-based method for reconstructing the local density and peculiar velocity fields, calculating a 64\% probability of the existence of the \ac{GA} as a unique watershed structure. These results align with the view that the \ac{GA} basin lies along a filamentary path feeding into more distant mass concentrations, and may not be a particularly significant region in its own right.

The classification of superclusters has evolved accordingly.
The Laniakea supercluster, which encompasses the \ac{GA} convergence point, was originally proposed by~\citet{Tully_2014} and is treated as one of several watershed basins in the local Universe by~\citet{Dupuy_2023}.
While initially characterised as a coherent basin, its membership and boundaries remain ambiguous; recent reconstructions indicate that it may in fact lie on the periphery of the more massive Shapley basin~\citep{Valade_2024}.
These revised interpretations cast the \ac{GA} not as a final destination of local flows but as a transition point along a larger trajectory.
As such, the classical image of a single dominant attractor gives way to a network of interacting flows and gravitational basins.
This also reflects the ongoing challenge of explaining the \ac{LG}'s velocity of $620~\kmsec$ relative to the \ac{CMB}~\citep{Planck_2020}---the original motivation for the \ac{GA} concept---which cannot be fully accounted for by attraction toward known overdensities alone.
In this context,~\citet{Hoffman_2017} introduced the complementary concept of ``repellers'', in particular proposing the ``Dipole Repeller'' as a large underdensity roughly opposite to Shapley also contributing to the \ac{LG}'s velocity.
The \ac{GA} concept thus carries an inherent ambiguity among three distinct physical questions: what mass distribution sources the \ac{LG}'s peculiar velocity in the \ac{CMB} frame (the original motivation of~\citealt{Bell_1988}), where the present-day velocity field converges (the streamline approach), and in what direction the \ac{LG} will actually move in the future.

These three definitions need not coincide.
Galaxies do not continue along their present-day streamlines indefinitely because these lines depict instantaneous velocities while the evolving mass distribution of the Universe causes velocities to change over time, so that actual dynamics may diverge significantly from the streamline picture. Indeed, in a dark-energy-dominated universe with low $\Omega_{\rm m}$, perturbation growth freezes out and structures decouple from the Hubble flow only if already bound, making the present velocity field encapsulated in streamlines a particularly misleading indicator of future trajectories.
An alternative approach is numerical action orbit modelling~\citep{Shaya_1995,Shaya_2017,Shaya_2022}, which computes past and future trajectories of galaxies within a \ac{LCDM} universe.

In this paper, we explicitly construct the \ac{GA} according to all three definitions using a new reconstruction of the local Universe based on the \acl{BORG} algorithm (\ac{BORG};~\citealt{Jasche_2013,Jasche_2019}), in particular the \texttt{Manticore-Local} implementation~\citep{McAlpine_2025}, which infers the initial and final density fields of the local ${\sim} 200~\Mpch$ by applying a sophisticated Bayesian forward model to the number density of galaxies in the \TWOMPP catalogue~\citep{Lavaux_2011}.
The velocity field derived therefrom has been shown to surpass other reconstructions by a substantial margin in accuracy~\citep{Stiskalek_2025_VFO,McAlpine_2025} and is therefore expected to afford a corresponding gain in \ac{GA} characterisation accuracy.
We apply the streamline method to these data, inferring the properties of the \ac{GA} and also investigating its smoothing-scale dependence, which we find to be considerable~(e.g.~\citealt{Dupuy_2019,Dupuy_2020}).
We also propose a new \emph{dynamical} definition of the \ac{GA} which, instead of assessing streamline convergence at fixed present time, explores the full trajectories of objects over cosmic history and into the future.
Zooming in further, the high-resolution \ac{BORG}-based reconstructions of the \ac{LG} neighbourhood by~\citet{Wempe_2024,Wempe_2025} are particularly well suited to studying the future evolution of the Milky Way.

The structure of the paper is as follows.
In Section~\ref{sec:data} we describe the \texttt{BORG} reconstructions that form the basis of our study and the catalogue of clusters that we investigate in a \ac{GA} context.
In Section~\ref{sec:method} we present our methodology, both for streamline convergence (Section~\ref{sec:streamlines}) and our new dynamical definition involving simulations evolved beyond the present time (Section~\ref{sec:future_sims}).
In Section~\ref{sec:results} we distinguish three definitions of the \ac{GA}, and present the results of our streamline and future dynamics analyses.
Section~\ref{sec:conclusion} concludes.
All logarithms are base-10.

\section{Data}\label{sec:data}

In Section~\ref{sec:local_universe_model}, we describe the \Manti digital twin suite, which provides a probabilistic reconstruction of the local Universe.
In Section~\ref{sec:random_sims}, we describe a set of random \ac{LCDM} simulations used to quantify the reconstruction volume necessary to capture the \ac{LG} motion.
In Section~\ref{sec:cluster_catalogue}, we present a compiled galaxy cluster catalogue used to identify observed structures in the \ac{GA} region.

\subsection{Digital twins of the local Universe}\label{sec:local_universe_model}

We employ the \Manti suite of digital twins,\footnote{Such digital twins are traditionally referred to as constrained simulations, however in our context a more precise name would be ``data-constrained posterior simulations'', highlighting that the initial conditions are derived from a full Bayesian forward model and include explicit uncertainty quantification.}
which provides a high-fidelity reconstruction of the local Universe out to approximately $200~\Mpc$ from the Milky Way~\citep{McAlpine_2025}.
\Manti is based on the \ac{BORG} algorithm~\citep{Jasche_2013,Jasche_2019} applied to the \TWOMPP galaxy catalogue~\citep{Lavaux_2011} and constitutes the latest \ac{BORG}-based reconstruction of the \TWOMPP volume.
\ac{BORG} produces a posterior distribution of initial density fields at $z = 1000$ on a $256^3$ grid centred on the Milky Way in a $681~\Mpch$ box by forward-modelling structure formation with a gravity solver, incorporating redshift-space distortions, selection effects, and galaxy biasing, and comparing the resulting redshift-space galaxy distribution to observations via a generalised Poisson likelihood~\citep{Jasche_2015,Lavaux_2016,Leclercq_2017,Lavaux_2019,Porqueres_2019,Stopyra_2023}.
Independent \ac{BORG} posterior samples are post-processed by augmenting the inferred $256^3$ initial conditions with random small-scale modes and then evolved to $z = 0$ with an $N$-body simulation, yielding the probabilistic ensemble of digital-twin realisations consistent with observed large-scale structure.

The \Manti suite comprises $80$ posterior samples resimulated with \texttt{SWIFT}~\citep{Schaller_2024}.
The simulations adopt a $681~\Mpch$ box centred on the Milky Way with uniform resolution on a $1024^3$ mesh, yielding an initial condition spatial resolution of $0.67~\Mpch$ and a particle mass of $2.4 \times 10^{10}~\Msunh$.
The cosmological parameters are drawn from the Dark Energy Survey Year 3 analysis~\citep{DES_Y3}: $h{=}0.681$, $\Omega_{\rm m}{=}0.306$, $\Omega_{\rm b}{=}0.0486$, $\sigma_8{=}0.807$, and $n_{\rm s}{=}0.967$, assuming the \ac{LCDM} model.
\citet{McAlpine_2025,McAlpine2025_ManticoreCatalogue} demonstrated that \Manti exhibits excellent agreement with observed cluster masses and positions, accurately reproduces the local velocity field, and matches \ac{LCDM} predictions for the power spectrum and halo mass function.

The \Manti suite provides density and velocity fields generated from the $z = 0$ particle distributions via smoothed particle hydrodynamics~\citep{Monaghan_1992,Colombi_2007}, with a minimum of 32 neighbours for the smoothing kernel.
These fields are sampled on a $256^3$ grid with voxel size $2.7~\Mpch$, sufficient for capturing the large-scale dynamics.
In addition to analysing these present-day fields, we also evolve the \Manti initial conditions beyond $z = 0$ to study the future dynamics of the local Universe by introducing the \ac{BPT} simulation suite (see Section~\ref{sec:future_sims}).

\subsection{Random simulations}\label{sec:random_sims}

To quantify the reconstruction volume necessary to capture all contributions to the \ac{LG} motion in the \ac{CMB} frame, we employ a set of random \ac{LCDM} simulations as a control sample.
We use the fiducial set of \texttt{Quijote} simulations~\citep{Quijote}, comprising $15{,}000$ $N$-body realisations in a $(1000~\Mpch)^3$ box with $512^3$ particles.
These were run with \texttt{Gadget-III}~\citep{Springel_2005} adopting a flat \ac{LCDM} cosmology consistent with the \textit{Planck} 2018 results~\citep{Planck_2020_cosmo}: $h = 0.6711$, $\Omega_{\rm m} = 0.3175$, $\Omega_{\rm b} = 0.049$, $\sigma_8 = 0.834$, and $n_{\rm s} = 0.9624$.
The cosmological parameters differ slightly from those adopted in \Manti, but we assume this difference is negligible for our comparative purposes.

We use the publicly available density fields at $z = 0$ sampled on a $256^3$ grid with voxel size $3.9~\Mpch$, constructed using a piecewise cubic spline (PCS) mass assignment scheme.
From these density fields, we compute the corresponding linear theory velocity field in Fourier space as~\citep{Peebles_1980}
\begin{equation}\label{eq:linear_velocity}
    \bm{v}(\bm{k}) = -i f H(a) a \frac{\delta(\bm{k}) \bm{k}}{k^2},
\end{equation}
where $f = \Omega_{\rm m}^{0.55}$ is the dimensionless growth rate~\citep{Bouchet_1995,Wang_1998}, $H(a)$ is the Hubble parameter, $a$ is the scale factor, $\delta(\bm{k})$ is the Fourier transform of the overdensity field, and $\bm{k}$ is the wave vector.
For each simulation, we compute the observer velocity at the box centre sourced by all matter, $\bm{V}_{\rm box}$, and by matter within radius $R$, $\bm{V}_{\rm inner}(R)$. To compute $\bm{V}_{\rm inner}(R)$, we mask the overdensity field to zero beyond $R$ before transforming to Fourier space.
Because the Fourier formulation implicitly assumes periodic boundary conditions, we pad the density field with empty cells by $50\%$ of the box size on each side before computing the velocity field, thereby suppressing spurious contributions from periodic replicas.
We compute both $\bm{V}_{\rm box}$ and $\bm{V}_{\rm inner}(R)$ for all $15{,}000$ simulations, sampling $R$ from $10$ to $300~\Mpch$ in $10~\Mpch$ increments.

\subsection{Galaxy cluster catalogue}\label{sec:cluster_catalogue}

To characterise the \ac{GA} and assess observed cluster membership, we adopt the cluster catalogue compiled by~\citet{McAlpine_2025} and supplement it with additional clusters in the \ac{GA} region queried from the NASA/IPAC Extragalactic Database\footnote{\url{https://ned.ipac.caltech.edu}}.
The resulting catalogue includes well-studied clusters such as Centaurus, Hydra, and Norma in the \ac{GA} region, along with surrounding structures including Virgo, Shapley, Perseus, and Coma.
Of particular interest is the Centaurus--Crux cluster filament, comprising Centaurus, Abell 3565, Abell S0753, and Abell 3574, which we discuss in detail in Section~\ref{sec:future_results}.
\cref{tab:cluster_catalogue} lists the cluster positions in Galactic coordinates and their recession velocities.

\begin{table}
    \centering
    \caption{Selection of galaxy clusters considered in this work, with a preference for clusters in the \ac{GA} region. Positions are given in Galactic coordinates and velocities are \ac{CMB} frame recession velocities. Coordinates are adopted from~\protect\citet{McAlpine_2025} or queried from the NED database. Multiple entries for Hercules and Shapley denote distinct clusters traditionally considered members of the respective superclusters.}
    \label{tab:cluster_catalogue}
    {\renewcommand{\arraystretch}{1.2}%
    \small
    \begin{tabular}{lccc}
    \hline
    Name & $\ell~[\deg]$ & $b~[\deg]$ & $cz_{\rm CMB}~[\kmsec]$ \\
    \hline
    Ursa Major & 144.57 & 65.5 & 1101 \\
    Fornax & 236.7 & $-53.0$ & 1332 \\
    Virgo & 283.8 & 74.4 & 1636 \\
    Centaurus & 302.4 & 21.6 & 3403 \\
    Hydra & 269.6 & 26.5 & 4058 \\
    Abell 3565 & 313.5 & 28.0 & 4120 \\
    Abell S0753 & 320.3 & 28.4 & 4200 \\
    Pavo II & 332.0 & $-24.0$ & 4200 \\
    Abell 3574 & 317.5 & 30.9 & 4942 \\
    Norma & 325.3 & $-7.1$ & 4955 \\
    Perseus (A426) & 150.6 & $-13.3$ & 4995 \\
    Leo & 235.1 & 73.0 & 6890 \\
    Abell 3581 & 323.1 & 32.9 & 7032 \\
    Coma & 58.1 & 88.0 & 7463 \\
    Hercules (A2199) & 62.9 & 43.7 & 9113 \\
    Abell 496 & 209.6 & $-36.5$ & 9849 \\
    Hercules (A2063) & 12.8 & 49.7 & 10634 \\
    Hercules (A2151) & 31.6 & 44.5 & 11024 \\
    Hercules (A2147) & 29.0 & 44.5 & 11072 \\
    Shapley (A3571) & 316.3 & 28.6 & 11965 \\
    Abell 548 & 230.3 & $-24.8$ & 12363 \\
    Abell 119 & 125.7 & $-64.1$ & 13004 \\
    Abell 1736 & 312.6 & 35.0 & 13823 \\
    Abell 1644 & 304.9 & 45.4 & 14448 \\
    Shapley (A3558) & 312.0 & 30.7 & 14784 \\
    Shapley (A3562) & 313.3 & 30.4 & 15065 \\
    \hline
    \end{tabular}}
    \vspace{1em}
\end{table}

\section{Methodology}\label{sec:method}

We adopt two complementary approaches to identify the \ac{GA}-like structures.
In Section~\ref{sec:streamlines}, we integrate streamlines of the present-day velocity field to identify convergence points and partition the velocity field into basins of attraction, following standard methods from the literature~\citep[e.g.][]{Dupuy_2019,Dupuy_2020,Dupuy_2025}.
In Section~\ref{sec:future_sims}, we evolve the \Manti initial conditions forward in time to determine the long-term gravitational fate of structures in the local Universe.

\subsection{Present-day velocity streamlines}\label{sec:streamlines}

Streamlines follow the scheme of~\citet{Dupuy_2019,Dupuy_2020,Dupuy_2025}: test particles are initialised at voxel centres and advanced with a second-order Runge-Kutta (RK2) integrator using an adaptive temporal step that enforces a fixed spatial advance.
We note that~\citet{Dupuy_2025} employ a higher-order Runge-Kutta integrator; we find this makes no substantial difference as we adopt small step sizes.
Velocities at particle positions are obtained via trilinear interpolation of the smoothed grid.
The RK2 update computes the velocity at the initial position, advances to an intermediate location, and evaluates the velocity at that half-step position (denoted $\bm{v}[\bm{x}]$ for velocity evaluated at position $\bm{x}$):
\begin{align}
    \bm{x}(t_{i+1/2}) &= \bm{x}(t_i) + \frac{1}{2} \bm{v}[\bm{x}(t_i)] \Delta t_i, \nonumber \\
    \bm{x}(t_{i+1}) &= \bm{x}(t_{i+1/2}) + \bm{v}[\bm{x}(t_{i+1/2})] \Delta t_{i+1/2}.
\end{align}
The temporal step maintains spatial increment $\Delta s$, such that
\begin{equation}
    \Delta t = \frac{\Delta s}{|\bm{v}| + \epsilon},
\end{equation}
with $\Delta s = 0.05 \Delta x$ for voxel size $\Delta x$ and $\epsilon = 10^{-6}~\kmsec$ to regularise vanishing velocities and avoid division by zero, yielding ${\sim}20$ integration steps per voxel.
We apply periodic boundary conditions, though all trajectories of interest remain in the central region of the box.
The streamline integration is similar to the dynamical definition of attractors (presented next), except that the velocity field is fixed to the present-time snapshot rather than evolved self-consistently under gravity, and the trajectories correspond to test particles advected through this field.
We assess convergence by monitoring particle displacement over a fixed fraction of the integration: the total number of steps is 25,000, and the monitoring window spans $10\%$ of them.
A streamline is converged when its displacement within the window falls below half a grid cell; with the choice of 25,000 steps we find that all particles meet this criterion.

Converged positions are binned to voxels and grouped into basins by merging all neighbouring (face-sharing) voxels.
Each component centroid is defined as the mean position of its member test particles, and the membership count measures the basin volume feeding that attractor.
The observer (i.e.\,Milky Way or \ac{LG}) convergence point is isolated by integrating a single streamline launched from the observer position, adopting the same convergence criterion.
We smooth the velocity field with a Gaussian kernel of standard deviation $\sigma_{\rm smooth}$, starting with no smoothing and then increasing the smoothing scale from $1$ to $16~\Mpch$ in steps of $1~\Mpch$, to isolate the large-scale flow while progressively suppressing small-scale fluctuations as $\sigma_{\rm smooth}$ is raised, and then follow streamlines over that smoothed field.
Across smoothing scales, we shall find that this trajectory terminates in Virgo, the \ac{cGA}, or Shapley, revealing how progressively stronger smoothing shifts the Milky Way streamline toward the larger-scale attractor.
We distinguish the GA \emph{convergence point} --- the actual convergence location of streamlines --- from the GA \emph{basin}, which is the set of all locations whose streamlines terminate at the GA point. Note that this distinction is sometimes confused in the literature; for example the Laniakea supercluster, which~\citet{Tully_2014,Dupuy_2023} treat as the basin associated with the GA, is simply the GA basin in our definition.

\subsection{Simulating future dynamics of the Local Universe}\label{sec:future_sims}

The streamline approach described in Section~\ref{sec:streamlines} identifies attractors directly from the present-time velocity field and exploits the long-range correlation of the velocity field to extrapolate the velocity field into the \ac{ZoA}.
While operationally useful for partitioning the flow into basins, this snapshot-based method has fundamental limitations: streamline convergence points depend on the chosen smoothing scale, do not necessarily coincide with physical structures (i.e.\ gravitationally bound systems), and do not describe the actual trajectories of test particles under gravitational evolution.
The instantaneous velocity field reflects the matter distribution at a single time slice and therefore cannot account for future mergers, the influence of dark energy, or the time-dependent strength of gravitational interactions.

A key advantage of \Manti is that it reconstructs not only the present-day density and velocity fields but also the underlying initial conditions and the full dynamical evolution in between. This enables us to move beyond snapshot-based inferences and follow the genuine structure formation dynamics. By evolving the \Manti initial conditions forward to a scale factor of $a = 10$, we allow gravitational interactions to unfold self-consistently and can identify which structures truly dominate the future evolution of the local Universe.
We evolve 50 \Manti realisations using the \texttt{Gadget4}~\citep{Gadget4} code with the same cosmology as \Manti.\footnote{The main \Manti suite contains 80 posterior samples; we use 50 (randomly chosen) in the \ac{BPT} for convenience. We verify that our results are not sensitive to this choice and that 50 posterior samples are sufficient.}
We choose $a = 10$ because structure formation effectively ceases as dark energy dominates the energy density (within the flat \ac{LCDM} cosmology assumed by \Manti; e.g.\,dynamical dark energy models would modify this timescale). We verify that the halo mass function no longer evolves significantly beyond $a = 2$.
Since we only require the large-scale dynamics, we adopt a low-resolution simulation with $256^3$ particles in a $681~\Mpch$ box, corresponding to a particle mass of approximately $1.6\times10^{12}~\Msunh$, sufficient to identify the cluster population in the local Universe.

We store particle snapshots at scale factors of 1, 2, 5, 10 (corresponding to ages of $13.8,\,24.8,\,40.4,\,52.4~{\rm Gyr}$) and identify dark matter haloes at each snapshot using the \texttt{Gadget4} built-in friends-of-friends (FoF) halo finder~\citep{Davis_1985_FoF} with a linking length of $0.2$ times the mean inter-particle separation.
We shall refer to this suite as the \Manti, \acf{BPT} suite.

\section{Results}\label{sec:results}

We now present results from both approaches to identifying the \ac{GA}.
In Section~\ref{sec:streamline_results}, we analyse streamlines of the present-day velocity field to locate the \ac{GA} and characterise its basin of attraction.
In Section~\ref{sec:future_results}, we track the gravitational evolution of structures in the local Universe to determine their long-term fate.
In Section~\ref{sec:lg_velocity}, we assess which scales contribute to the \ac{LG} velocity in the \ac{CMB} frame.

As discussed above, the \ac{GA} label mixes several related issues; to disentangle them, we distinguish three definitions:
\begin{enumerate}
    \item \textit{\acf{cGA}}: The large-scale velocity field structure identified historically from peculiar-velocity samples.~\citet{Dressler_1987} fitted a bulk flow (constant peculiar velocity), whereas~\citet{Bell_1988} fitted an attractor model in which peculiar velocity falls off as a power law of distance from the attractor centre. More recent reconstructions (e.g.\,\citealt{Dupuy_2025}) locate the \ac{cGA} through streamline convergence of the present-day velocity field.~\citet{Bell_1988} placed this near $(\ell,\,b) \approx (307,\,9)\,\deg$ at ${\sim}43~\Mpch$, with Norma later proposed as the \ac{GA} core at $(\ell,\,b) = (325.3,\,-7.1)\,\deg$ and $50~\Mpch$~\citep{Kraan-Korteweg_1996}.
    \item \textit{\acf{sGA}}: Either the true (model-dependent) observer streamline convergence point or its associated basin. The analyses referred to above identify the \ac{sGA} convergence point with the \ac{cGA}, although we find the \ac{sGA} to be significantly smoothing-dependent:
    at intermediate smoothing ($\sigma_{\rm smooth} \approx 3~\Mpch$) the \ac{sGA} convergence point coincides with the \ac{cGA}, but it transitions to Virgo at low smoothing and to Shapley at high smoothing. The \ac{cGA} inferred by our analysis is therefore the \ac{sGA} at such intermediate smoothing.
    \item \textit{\acf{dGA}}: The asymptotic future displacement direction of the Milky Way under full gravitational evolution. We determine this by evolving constrained initial conditions of the local Universe beyond present-time in the \ac{BPT} suite rather than integrating the frozen present-time velocity field as in the above definition. By evolving to $a = 10$, we identify which structures are gravitationally dominant and nearly bound to the \ac{LG}; dark energy prevents full gravitational collapse, so we cannot follow trajectories to $a \to \infty$, but the dominant gravitational influences are already apparent by this epoch.
\end{enumerate}

Note that while the \ac{sGA} and \ac{dGA} each have a basin associated with the convergence point, the \ac{cGA} is typically defined as the convergence point itself. We do not consider the basin associated with the \ac{dGA} (the region within which test particles all end up at the same place), so \ac{dGA} will refer specifically to the convergence point henceforth.

\subsection{Streamline analysis}\label{sec:streamline_results}

\begin{figure*}
    \centering
    \includegraphics[width=\textwidth]{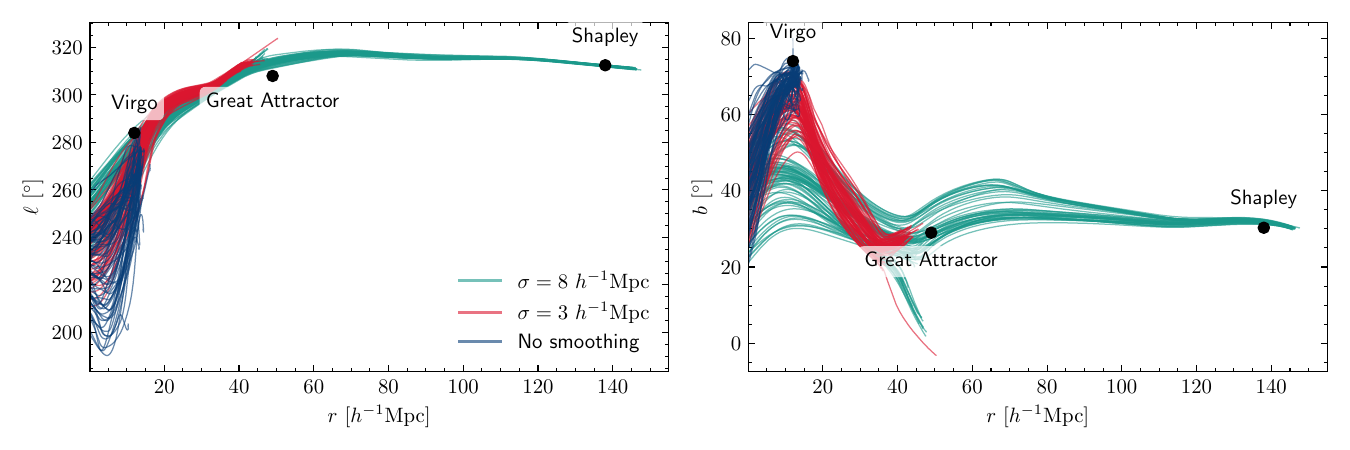}
    \caption{
        Milky Way streamline convergence points in Galactic coordinates for different smoothing scales applied to the \Manti velocity fields.
        Left and right panels show the Galactic longitude and latitude as functions of distance from the observer, respectively.
        Colours represent different smoothing scales; each line shows the streamline from a single \Manti realisation.
        Three discrete endpoints emerge: without smoothing, streamlines converge to Virgo; at $\sigma_{\rm smooth}=3~\Mpch$ to the \ac{cGA}; and at $\sigma_{\rm smooth}=8~\Mpch$ to Shapley.
        Black points indicate the approximate positions of these structures, with the \acl{cGA} position adopted from~\citet{Dupuy_2025}.
        }
    \label{fig:mw_streamlines_smoothing}
\end{figure*}

We first examine the Milky Way streamline and its smoothing scale-dependence to localise the \ac{sGA} within the \Manti realisations.
Following~\citet{Dupuy_2019,Dupuy_2020}, we identify the \ac{sGA} convergence point by tracing a streamline from the observer (Milky Way or \ac{LG}) position to its terminus.
The streamlines shown in~\cref{fig:mw_streamlines_smoothing} display three convergence regimes: Virgo when $\lesssim2~\Mpch$ smoothing is applied, the \ac{cGA} for $2~\Mpch \lesssim \sigma_{\rm smooth} \lesssim 4~\Mpch$, and Shapley at high smoothing values.
This occurs not because the location of the ``drain'', i.e.\ the convergence point, is sensitive to the smoothing, but rather because the ``mountain ranges'' that separate watersheds are sensitive to smoothing: when the smoothing is sufficiently large, separate basins are merged together.
For $\sigma_{\rm smooth}=2~\Mpch$, $30$ of $80$ Milky Way streamlines still terminate in Virgo, whereas all streamlines at either $\sigma_{\rm smooth}=3$ or $4~\Mpch$ end at the \ac{cGA}.
At $\sigma_{\rm smooth} = 8~\Mpch$, approximately $20\%$ of the streamlines do not reach Shapley but instead typically terminate at ${\sim}50~\Mpch$ near $b \approx 0\,\deg$, i.e.\ in the \ac{ZoA}. Given the incomplete knowledge of the large-scale structure, probabilistic treatments such as \ac{BORG} are required to determine the plausible range of realisations compatible with the observed data.

Following streamlines from the observer position at smoothing scales of $2,\,3$ and $4~\Mpch$, we identify the \ac{cGA} convergence point in each \Manti realisation.
\cref{tab:ga_positions} presents the inferred positions in Galactic coordinates $(r,\,\ell,\,b)$.
At $\sigma_{\rm smooth} = 3~\Mpch$, the inferred \ac{cGA} convergence point lies at distance $41^{+2}_{-5}~\Mpch$ and $(\ell,\,b) = (313^{+1}_{-5},\,27^{+2}_{-4})\,\deg$, close to Abell 3565 (see~\cref{tab:cluster_catalogue}).
Abell 3565 is part of a filamentary structure extending through Centaurus, Abell S0753 and Abell 3574, which are members of the Hydra--Centaurus Supercluster.
This nexus of flows has been identified previously: the cluster grouping including Abell~3565, S0753, and Abell~3574 was discussed as part of the ``four clusters'' region by~\citet{Courtois_2013}, and it marks the terminus of flows in the Laniakea basin defined by~\citet{Tully_2014}; see also the flow lines computed by~\citet{Shaya_2022}.
This is in
good agreement with the recent result of~\citet{Dupuy_2025}, who reported $(\ell,\,b) = (308.4 \pm 2.4\,\deg,\,29.0 \pm 1.9\,\deg)$ at a distance of $50 \pm 4~\Mpch$.

\begin{figure}
    \centering
    \includegraphics[width=\columnwidth]{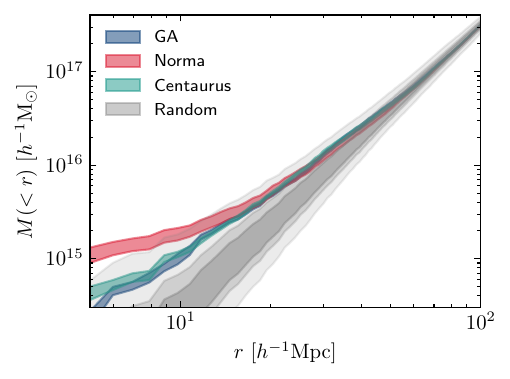}
    \caption{
        Enclosed mass profile around the \Manti-inferred \acl{cGA} position at $(\ell,\,b) = (313^{+1}_{-5},\,27^{+2}_{-4})\,\deg$ and distance $r=41^{+2}_{-5}~\Mpch$ identified from the Milky Way streamline at $\sigma_{\rm smooth}=3~\Mpch$, averaged over 80 \Manti realisations.
        For comparison, we also show profiles centred on Norma, Centaurus, and random positions.
        The profile exhibits a $2-3\sigma$ overdensity out to $20~\Mpch$, confirming a significant mass concentration at this location.}
    \label{fig:ga_enclosed_mass}
\end{figure}

In~\cref{fig:ga_enclosed_mass}, we show the enclosed mass profile around the \ac{cGA} convergence point at $\sigma_{\rm smooth} = 3~\Mpch$, i.e. the \ac{cGA} inferred by \Manti.
The enclosed mass is obtained by summing voxel masses within radius $r$ of the convergence point position, averaged over the \Manti realisations.
The profile shows a $2-3\sigma$ overdensity out to about $20~\Mpch$.
However, when we compare the enclosed mass profile directly with the nearby Centaurus cluster, we find that the \ac{cGA} overdensity is similar.
Furthermore, we compare the enclosed mass to that of Norma, which has been reported to be at the centre of the \ac{cGA}~\citep{Woudt_2000,Woudt_2008} but which we shall find not to belong to the \ac{sGA} basin.
The enclosed mass profile around Norma is significantly larger up to about $10~\Mpch$, reflecting not only its higher mass compared to Centaurus but also that both Norma and Centaurus pointings are centred directly on the haloes representing these clusters, whereas the \ac{cGA} convergence point does not necessarily correspond to the centre of a massive halo.

\begin{table}
    \centering
    \caption{Streamline Great Attractor convergence coordinates of the Milky Way streamline across Gaussian smoothing scales.}
    \label{tab:ga_positions}
    {\renewcommand{\arraystretch}{1.2}%
    \begin{tabular}{lccc}
    $\sigma_{\rm smooth}~[\Mpch]$ & $r~[\Mpch]$ & $\ell~[\deg]$ & $b~[\deg]$ \\
    \hline
    $2$ & $41.0_{-6.4}^{+2.0}$ & $313.0_{-7.8}^{+1.4}$ & $27.3_{-4.1}^{+1.4}$ \\
    $3$ & $41.3_{-4.7}^{+2.0}$ & $313.0_{-4.8}^{+1.1}$ & $27.4_{-4.3}^{+1.8}$ \\
    $4$ & $41.8_{-3.5}^{+1.5}$ & $312.0_{-2.3}^{+1.3}$ & $26.7_{-4.6}^{+2.5}$ \\
    \hline
    \end{tabular}}
\end{table}

\begin{figure*}
    \centering
    \includegraphics[width=\textwidth]{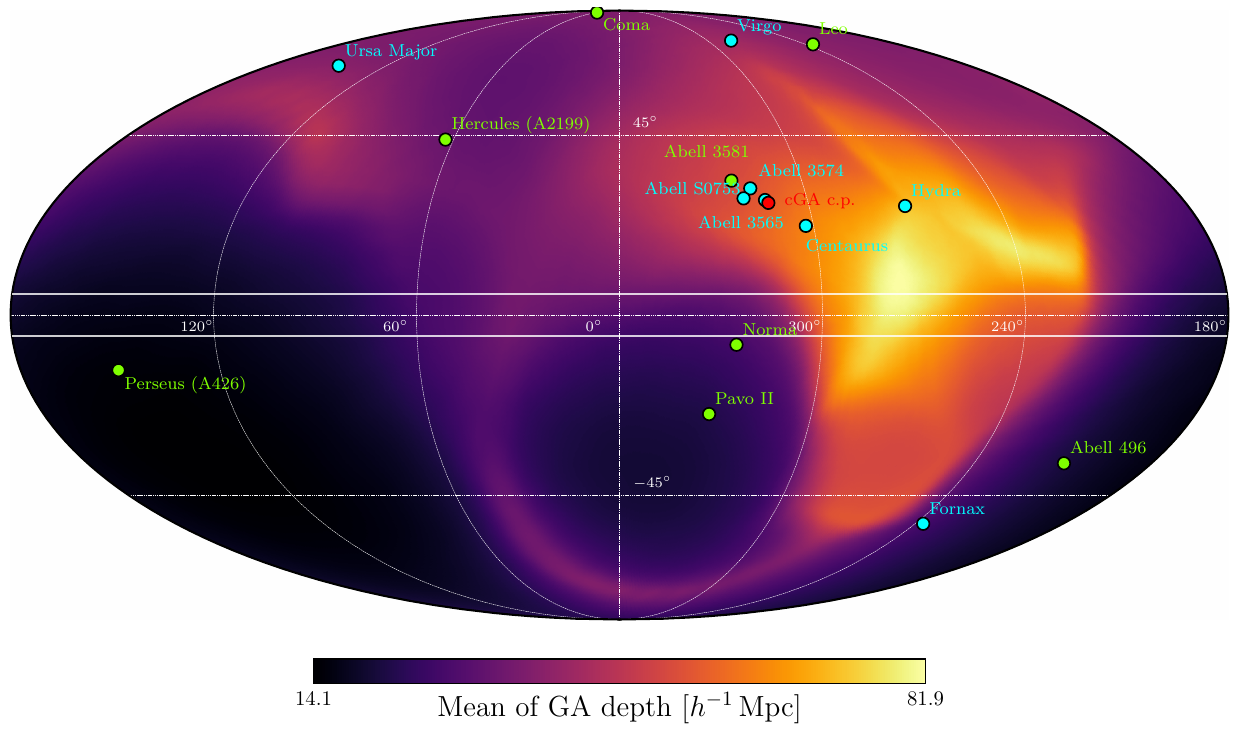}
    \caption{
        Depth of the inferred \acl{cGA} basin as a function of sky position in Galactic coordinates, averaged over 80 \Manti realisations.
        The basin is defined by selecting all voxels whose streamlines converge to the \ac{cGA} position identified from the Milky Way streamline at $\sigma_{\rm smooth}=3~\Mpch$, and ``cGA c.p.'' indicates its convergence point.
        For each sky direction, the depth is computed as the distance to the farthest voxel whose streamline converges to the \ac{cGA} position.
        The basin is highly elongated, reaching six times farther toward the convergence point than away from it.
        Clusters that belong to the \ac{cGA} basin in $>50\%$ of the realisations are plotted in cyan,
        whilst other clusters from~\cref{tab:cluster_catalogue} within $100~\Mpch$ are shown in green.
        The inferred \ac{cGA} convergence point is also indicated in cyan; the \ac{ZoA} (white lines at $b = \pm 5\,\deg$) is shown for reference.
        Dashed white lines indicate meridians every $60\,\deg$ and parallels at $\pm 45\,\deg$.
        }
    \label{fig:ga_depth_sky}
\end{figure*}

We now characterise the basin associated with the \ac{cGA} convergence point by selecting all voxels whose streamlines converge to the identified \ac{sGA} position and computing the total enclosed mass and equivalent spherical radius.
We find the total mass of all such voxels to be $\log (M/(\Msunh))\!=\!16.4^{+0.1}_{-0.1}$, and the corresponding equivalent spherical radius is $R = 42^{+3}_{-1}~\Mpch$, though the basin is highly non-spherical (see~\cref{fig:ga_depth_sky}).
A sphere of this radius at the mean matter density ($\bar{\rho}_m = \Omega_{\rm m} \rho_{\rm c}$) would enclose the same mass, indicating that the over- and under-dense regions within the basin largely cancel.

Since the observer streamline converges to the \ac{cGA}, we are located within this basin.
In~\cref{fig:ga_depth_sky} we show the depth of the inferred \ac{cGA} basin as a function of sky position in Galactic coordinates, computed by finding the distance to the farthest voxel whose streamline converges to the attractor for each sky direction.
The basin extends to ${\sim}80~\Mpch$ in the direction of the convergence point (near the Centaurus and Hydra clusters) and only ${\sim}13~\Mpch$ in the opposite direction.
A narrow protrusion of the \ac{cGA} basin extends through the southern Galactic hemisphere from Fornax back toward the north, likely threading between the neighbouring Perseus--Pisces and Norma basins.
Appendix~\ref{sec:appendix_ga_figs} presents the standard deviation of the basin depth across the 80 \Manti realisations, together with the sky-projected mean density field.
Across the 80 \Manti realisations there is no tendency for another, comparably massive system to appear directly behind the \ac{ZoA}.
However, when testing \Manti against the CosmicFlows-4 Tully--Fisher sample~\citep{Kourkchi_2020}, a residual external velocity $V_{\rm ext} = 90 \pm 8~\kmsec$ toward $(\ell,\,b) = (310 \pm 6,\,-5 \pm 4)\,\deg$ remains, roughly aligned with the \ac{ZoA}~\citep{McAlpine_2025}.
This residual could indicate observational systematics, an unresolved structure within the reconstruction volume (e.g.\ in the \ac{ZoA}), or the influence of mass beyond the ${\sim}200~\Mpc$ covered by the \TWOMPP volume. We leave detailed investigation of this residual to future work with the \ac{BORG} framework jointly analysing peculiar-velocity data in order to constrain the presence of massive structures behind the \ac{ZoA}.
Velocity-field-based reconstructions employing forward modelling of peculiar velocities (e.g.~\citealt{Valade_2024}) may be better at probing the mass distribution in obscured regions, since peculiar velocities are sensitive to larger scales. While~\citet{Hollinger_2024} quantified how rapidly linear-theory predictions based on \TWOMPP redshift data degrade in the \ac{ZoA}, the extent to which \ac{BORG} mitigates this degradation has not yet been tested, nor has a systematic comparison with velocity-field-based methods in these regions been carried out. We note, however, that because \ac{BORG} infers a continuous \ac{LCDM} realisation of the density field, it must smoothly interpolate between regions where data are available, placing implicit constraints on the \ac{ZoA}.
Our inferred \ac{cGA} position already coincides well with that reported by~\citet{Dupuy_2025}, who jointly analysed both redshift-space galaxy counts and peculiar velocities.

Of the galaxy clusters in our catalogue (Section~\ref{sec:cluster_catalogue}), eight are members of the \ac{cGA} basin in more than $50\%$ of the \Manti realisations, as determined by whether their reported positions lie within the basin.
Hydra, Centaurus, Virgo, Abell 3565, Abell S0753, Fornax, and Ursa Major belong to the basin in all \Manti realisations, whilst Abell 3574 is a member in $97\%$ of the realisations.
In contrast to previous studies, we find Norma (and Pavo~II) to lie outside of the \ac{cGA} basin.
At low Galactic latitude, the Ophiuchus cluster (NeVe~1; $(\ell,\,b) \approx (1.3,\,9.7)\,\deg$, $cz_{\rm CMB} \approx 8550~\kmsec$) is recovered by \Manti as a halo of mass $\log (M_{200c}/(\Msunh)) = 14.5^{+0.2}_{-0.3}$ at $(\ell,\,b) = (3.2^{+1.3}_{-1.8},\,10.6^{+1.0}_{-0.6})\,\deg$ and $cz = 8810^{+140}_{-170}~\kmsec$, though at $|b| \approx 10\,\deg$ it lies outside the effective \TWOMPP Galactic mask and is largely present in the input data.

\subsection{Future dynamics analysis}\label{sec:future_results}

We previously identified the \ac{GA} as a convergence point of the present-day velocity field, but this construction reflects the instantaneous matter distribution, not the future dynamics.
To determine which structures dominate the future displacement of the \ac{LG} neighbourhood, we analyse the \ac{BPT} suite of $N$-body simulations evolved from the \Manti initial conditions to scale factor 10 (see Section~\ref{sec:future_sims}).

\begin{figure*}
    \centering
    \textbf{Dynamical Great Attractor: Future displacement towards Virgo}\\[1ex]
    \includegraphics[width=\textwidth]{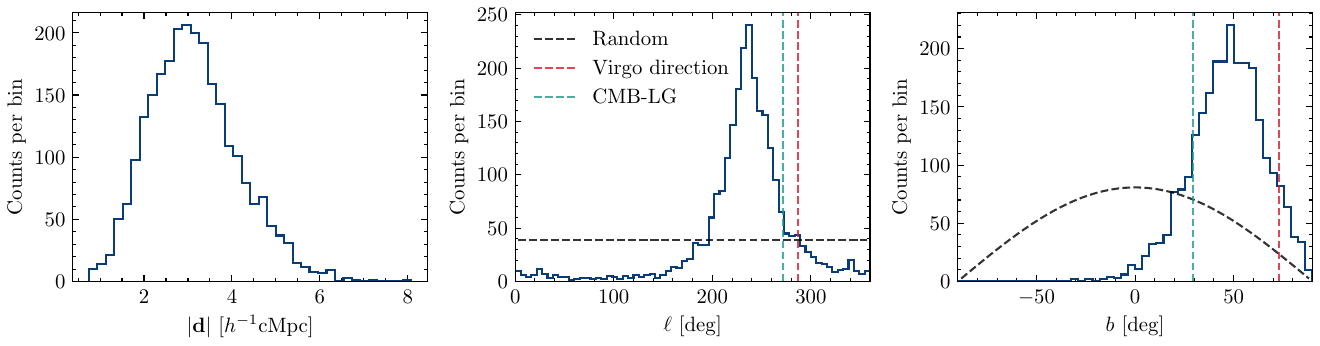}
    \caption{
        Displacement vectors of particles within $R = 5~\cMpch$ of the observer position (centre of the box), tracked from $a = 1$ to $a = 10$, shown in Galactic coordinates (magnitude, longitude, and latitude, respectively) stacked over the simulations.
        Particles typically move towards the Virgo cluster but fall short of reaching it; Virgo's distance from the box centre at $a = 10$ is approximately $14~\cMpch$.
        Vertical lines mark Virgo at $(\ell,\,b) = (287.0,\,73.2)\,\deg$ (red) and the \ac{LG} \ac{CMB}-frame velocity at $(\ell,\,b) = (271.9,\,29.6)\,\deg$ (cyan) as reported by \textit{Planck}~\citep{Planck_2018}.
        The $\ell$ and $b$ panels show the expected isotropic distribution as black dashed lines.}
    \label{fig:observer_particle_displacements}
\end{figure*}

We select particles within radius $R = 5~\cMpch$ of the observer position and track their evolution to $a = 10$.
Choosing a larger radius would lead to enclosing particles close to the Virgo cluster, whose distance from the origin is ${\sim}13~\Mpch$.
In~\cref{fig:observer_particle_displacements} we show the particle displacement magnitudes and directions in Galactic coordinates.
Particles move on average towards the Virgo region but typically travel only $3 \pm 1~\cMpch$, well short of the ${\sim}14~\cMpch$ separating Virgo from the box centre at $a=10$.
The displacement direction is $(\ell,\,b) = (232 \pm 51,\,47 \pm 19)\,\deg$, broadly aligned with Virgo and with the \ac{LG} velocity in the \ac{CMB} frame~\citep{Planck_2018}, though shifted to somewhat lower longitude and latitude, reflecting the complex dynamics in this region.
None of these particles become gravitationally bound to the Virgo halo by $a = 10$.
Across the \Manti realisations, the present-time observer velocity has mean $|v| = 457 \pm 56~\kmsec$ at $(\ell,\,b) = (244 \pm 12,\,39 \pm 10)\,\deg$, offset to lower longitude and latitude than the \ac{LG} \ac{CMB}-frame velocity of ${\sim}620~\kmsec$ towards $(272,\,30)\,\deg$~\citep{Planck_2018}, consistent with the influence of Perseus and Fornax.
However, the residual scatter of ${\sim}150~\kmsec$ from unresolved small-scale modes (below the \ac{BORG} constraint scale) at galaxy-group scales complicates any stringent comparison between the \ac{BORG} observer velocity and the \ac{LG} \ac{CMB}-frame velocity~\citep{Stiskalek_2025_VFO}.
We consider the \ac{LG} velocity further in Section~\ref{sec:lg_velocity}.
We verify that this conclusion holds even when evolving the simulations to $a = 100$.
Examining the future displacement of the Virgo cluster itself in \Manti, we find that at present its distance is $13^{+1}_{-1}~\cMpch$ and its direction is $(\ell,\,b) = (287^{+4}_{-2},\,73^{+1}_{-1})\,\deg$.
By $a = 10$, Virgo moves to $r = 14^{+1}_{-1}~\cMpch$ and $(\ell,\,b) = (284^{+3}_{-4},\,67^{+1}_{-1})\,\deg$, suggesting a mild outward displacement in the direction of $(\ell,\,b) = (277^{+11}_{-14},\,19^{+11}_{-14})\,\deg$.

The streamlines from the Milky Way identify the \ac{cGA} convergence point near the Abell 3565 cluster, which is represented in \Manti at $a=1$ by a halo of mass $\log (M_{200c}/(\Msunh))\!=\!13.8^{+0.1}_{-0.2}$, although it appears in only $70\%$ of the realisations.
The other massive cluster close to this position is Abell S0753, with mass of $\log (M_{200c}/(\Msunh))\!=\!14.0^{+0.1}_{-0.1}$, which is present in all \Manti realisations.
In realisations where both haloes are present at $a=1$, they typically merge before $a=10$, leading to a halo near the \ac{cGA} position inferred from the streamlines, close to the present-time position of Abell~3565.
When Abell~3565 is absent, the Abell~S0753 halo generally migrates towards the \ac{cGA} convergence point by $a=10$.

\cref{tab:manticore_clusters} summarises the positions of the observed clusters near the \ac{cGA} convergence point and their corresponding \Manti present-time haloes, together with their halo masses and streamline-basin memberships.
There is excellent agreement with the observed positions of all clusters.
The other massive cluster near the \ac{cGA} position is Centaurus, which in \Manti has a mass of $\log (M_{200c}/(\Msunh))\!=\!14.3^{+0.1}_{-0.1}$.
By $a=10$ it grows to $\log (M_{200c}/(\Msunh))\!=\!14.4^{+0.1}_{-0.1}$ and undergoes only mild displacement, reaching $r = 35^{+1}_{-1}~\cMpch$ and $(\ell,\,b) = (306^{+1}_{-1},\,24^{+1}_{-1})\,\deg$.

This indicates that Virgo, not the previously identified \ac{cGA} convergence point, is the dominant (bound) structure determining the future \ac{LG} displacement, with secondary contributions from other nearby massive clusters as quantified in~\cref{tab:force_contributions} and discussed below.
Although there is a clear convergence point of the Milky Way streamlines, our dynamical analysis shows that the \ac{LG}'s future displacement is directed towards Virgo as the \ac{dGA}, not this intermediate-scale structure.
In fact, this result agrees with the streamline analysis in Section~\ref{sec:streamline_results}: when no smoothing is applied to the velocity field, the observer streamline converges to Virgo.

\Manti identifies Norma and Perseus as the most massive systems within $6000~\kmsec$, with masses of $\log (M_{200c}/(\Msunh))\!=\!14.8^{+0.1}_{-0.1}$ and $\log (M_{200c}/(\Msunh))\!=\!14.9^{+0.1}_{-0.1}$, respectively (see \cref{tab:manticore_clusters}).
Neither cluster is part of the \ac{sGA} basin in our streamline analysis (Section~\ref{sec:streamline_results}), likely because each forms its own basin of attraction.
The next most massive cluster is Coma, although it lies farther away at ${\sim}74~\Mpch$.

\begin{table*}
    \centering
    \caption{
        Comparison of observed positions (\cref{tab:cluster_catalogue}) with \Manti halo properties at $a = 1$ for selected galaxy clusters in the \ac{GA} region.
        Coordinates are given in Galactic coordinates and clusters are ordered by their observed redshift.}
    \label{tab:manticore_clusters}
    {\renewcommand{\arraystretch}{1.2}%
    \small
    \setlength{\tabcolsep}{4pt}
    \begin{tabular}{lccc cccccc}
    \hline
    & \multicolumn{3}{c}{Observed} & & \multicolumn{5}{c}{\Manti} \\
    \cline{2-4} \cline{6-10}
    Name & $\ell~[\deg]$ & $b~[\deg]$ & $cz_{\rm CMB}~[\kmsec]$ & & $\ell~[\deg]$ & $b~[\deg]$ & $r~[\cMpch]$ & $\log M_{200c}~[\Msunh]$ & \ac{cGA} member \\
    \hline
    Inferred \ac{cGA} & -- & -- & -- & & $313.0^{+1.1}_{-4.8}$ & $27.4^{+1.8}_{-4.3}$ & $41.3^{+2.0}_{-4.7}$ & -- & -- \\
    Ursa Major & $144.6$ & $65.5$ & $1101$ & & $147.3^{+8.9}_{-6.3}$ & $70.7^{+2.7}_{-9.2}$ & $11.8^{+1.8}_{-2.8}$ & $13.6^{+0.1}_{-0.3}$ & $100\%$ \\
    Fornax & $236.7$ & $-53.0$ & $1332$ & & $235.6^{+6.5}_{-5.1}$ & $-53.5^{+1.9}_{-1.2}$ & $13.5^{+1.3}_{-1.1}$ & $13.8^{+0.2}_{-0.2}$ & $100\%$ \\
    Virgo & $283.8$ & $74.4$ & $1636$ & & $287.0^{+4.2}_{-1.5}$ & $73.2^{+1.4}_{-1.4}$ & $13.0^{+0.9}_{-0.5}$ & $14.4^{+0.1}_{-0.1}$ & $100\%$ \\
    Centaurus & $302.4$ & $21.6$ & $3403$ & & $303.0^{+0.4}_{-0.7}$ & $21.7^{+0.6}_{-0.5}$ & $34.1^{+0.8}_{-0.8}$ & $14.3^{+0.1}_{-0.1}$ & $100\%$ \\
    Hydra & $269.6$ & $26.5$ & $4058$ & & $269.6^{+0.3}_{-0.3}$ & $26.5^{+0.4}_{-0.4}$ & $43.5^{+1.1}_{-1.2}$ & $14.3^{+0.1}_{-0.1}$ & $100\%$ \\
    Abell 3565 & $313.5$ & $28.0$ & $4120$ & & $313.5^{+1.4}_{-1.1}$ & $29.0^{+0.9}_{-0.7}$ & $40.8^{+1.6}_{-1.0}$ & $13.8^{+0.1}_{-0.2}$ & $100\%$ \\
    Abell S0753 & $320.3$ & $28.4$ & $4200$ & & $319.5^{+0.6}_{-1.6}$ & $26.5^{+1.0}_{-0.5}$ & $42.9^{+1.3}_{-1.0}$ & $14.0^{+0.1}_{-0.1}$ & $100\%$ \\
    Abell 3574 & $317.5$ & $30.9$ & $4942$ & & $317.6^{+0.6}_{-0.5}$ & $31.5^{+0.4}_{-0.4}$ & $50.6^{+1.4}_{-1.6}$ & $13.9^{+0.2}_{-0.2}$ & $97\%$ \\
    Norma & $325.3$ & $-7.1$ & $4955$ & & $325.3^{+0.4}_{-0.3}$ & $-6.3^{+0.4}_{-0.4}$ & $51.0^{+1.1}_{-0.8}$ & $14.8^{+0.1}_{-0.1}$ & No \\
    Perseus & $150.6$ & $-13.3$ & $4995$ & & $150.3^{+0.4}_{-0.4}$ & $-13.5^{+0.3}_{-0.2}$ & $51.5^{+1.1}_{-2.2}$ & $14.9^{+0.1}_{-0.1}$ & No \\
    \hline
    \end{tabular}}
\end{table*}

Finally, we assess the relative influence of nearby haloes using a simple illustrative model in which the Newtonian gravitational force on an observer at the box centre scales as $F \propto M_{200c}/r^{2}$.
Importantly, individual clusters contribute only ${\sim}10~\kmsec$ to peculiar velocities; the bulk arises from integration over supercluster-scale matter distributions. Summing the contributions from bound haloes therefore falls well short of the observed ${\sim}600~\kmsec$, and this analysis serves only to illustrate the relative importance of nearby structures rather than to account for the full peculiar velocity budget; integration over all matter within a given radius is performed in Section~\ref{sec:lg_velocity}.
For each realisation we rank forces from all haloes on the observer, retain the top 100 contributors, and then stack these across realisations using the halo-association scheme of~\citet{McAlpine2025_ManticoreCatalogue} to identify persistent contributors across the posterior.
\Cref{tab:force_contributions} lists the nine largest contributions from haloes within $155~\Mpch$, normalised to the strongest contributor and interpreted as the principal drivers of the future \ac{dGA} direction.
The dominant contribution arises from haloes representing Virgo, followed by Perseus at roughly $22\%$ of Virgo's influence, Fornax at a similar level (present in $88\%$ of the realisations), Norma at about $17\%$, and Centaurus at about $13\%$.
In approximately $25\%$ of the realisations, a halo representing the nearby Ursa Major cluster~\citep{Trentham_2001} appears and exerts a force comparable to that of Perseus, though with mass of $\log (M_{200c}/(\Msunh))\!\approx\!13.5$ it sits near the resolution limit of our \ac{BPT} suite (but not of the higher-resolution \Manti main suite).

\begin{table}
    \centering
    \caption{Most influential haloes contributing to the gravitational force on the observer, ranked by $F \propto M_{200c}/r^{2}$ and normalised to the strongest contributor. Positions and distances are averaged over the \Manti realisations in which each halo is present; Count is the number of realisations (out of 50) containing the halo.}
    \label{tab:force_contributions}
    {\renewcommand{\arraystretch}{1.2}%
    \small
    \begin{tabular}{lccc}
    \hline
    Name & Count & $r~[\cMpch]$ & $F/F_{\max}$ \\
    \hline
    Virgo & 50 & 13.18 & 1.0 \\
    Perseus & 50 & 51.18 & 0.22 \\
    Fornax & 44 & 14.45 & 0.22 \\
    Ursa Major & 13 & 12.61 & 0.17 \\
    Norma & 50 & 51.15 & 0.17 \\
    Centaurus & 47 & 34.09 & 0.13 \\
    Coma & 49 & 72.20 & 0.10 \\
    Pavo II & 38 & 44.48 & 0.10 \\
    Antlia & 21 & 30.96 & 0.10 \\
    \hline
    \end{tabular}}
\end{table}

\subsection{Explaining the Local Group velocity}\label{sec:lg_velocity}

In \Manti, the present-day observer velocity has mean amplitude $|v| = 457 \pm 56~\kmsec$ directed towards $(\ell,\,b) = (244 \pm 12,\,39 \pm 10)\,\deg$, including the residual external velocity $V_{\rm ext} = 90 \pm 8~\kmsec$ inferred by comparing \Manti to the CosmicFlows-4 Tully--Fisher sample~\citep{Kourkchi_2020}.
This is both lower in amplitude and offset in direction from the \ac{LG} velocity in the \ac{CMB} frame of ${\sim}620~\kmsec$ towards $(272,\,30)\,\deg$~\citep{Planck_2018}.

Since~\citet{Bell_1988} originally invoked the \ac{GA} to explain the \ac{LG} velocity in the \ac{CMB} frame, we now assess which scales contribute to the \ac{LG} velocity within \Manti, following~\citet{Lavaux_2010}.
\citet{Lavaux_2010} reconstructed the local velocity field using the 2MASS galaxy catalogue and found that structures within $40~\Mpch$ of the \ac{LG} contribute approximately half the \ac{CMB} dipole amplitude.
They found no convergence in either amplitude or sky direction even at scales of $150~\Mpch$; mass within this distance yields a dipole of $500 \pm 100~\kmsec$, offset in direction by $40$--$50\,\deg$.
These findings are consistent with \Manti, which similarly exhibits a lower \ac{LG} dipole amplitude and a directional offset.
Using linear theory applied to the \TWOMPP compilation,~\citet{Carrick_2015} predicted a \ac{LG} velocity of $540 \pm 40~\kmsec$ towards $(\ell,\,b) = (268 \pm 4,\,38 \pm 6)\,\deg$, only $10\,\deg$ from the \ac{CMB} dipole direction. Nevertheless, they found a need for an external bulk flow of $159 \pm 23~\kmsec$ towards $(\ell,\,b) = (304 \pm 11,\,6 \pm 13)\,\deg$ at $5.1\sigma$ significance, indicating that sources beyond ${\sim}200~\Mpch$ (or in the \ac{ZoA}) still contribute appreciably.

\begin{figure*}
    \centering
    \includegraphics[width=\textwidth]{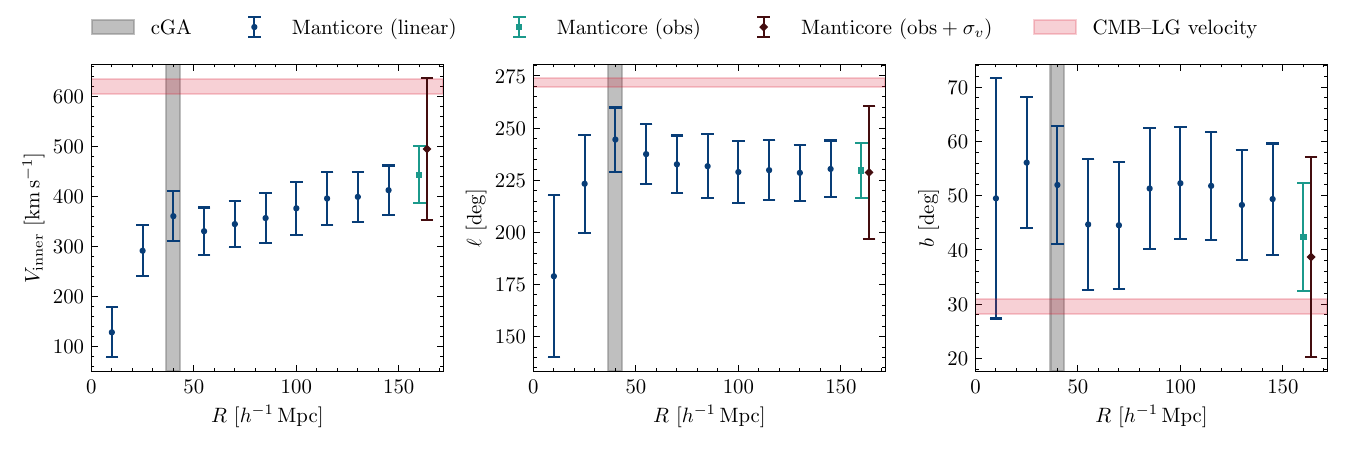}
    \caption{Observer velocity sourced by matter within radius $R$ in \Manti. \textit{Left}: Amplitude $|\bm{V}_{\rm inner}(R)|$. \textit{Centre}: Galactic longitude $\ell$ of $\bm{V}_{\rm inner}(R)$. \textit{Right}: Galactic latitude $b$ of $\bm{V}_{\rm inner}(R)$.
    ``\texttt{Manticore} (linear)'' shows the linear velocity sourced by matter within radius $R$, computed using~\cref{eq:linear_velocity}.
    ``\texttt{Manticore} (obs)'' denotes the observer velocity computed from the entire \Manti simulation box, including the external velocity dipole of $90 \pm 8~\kmsec$.
    ``\texttt{Manticore} (obs + $\sigma_v$)'' includes $150~\kmsec$ scatter to mimic non-linear contributions.
    Horizontal red bands indicate the observed \ac{LG} velocity in the \ac{CMB} frame~\citep{Planck_2018}.
    Vertical grey bands indicate the 16th--84th percentile on the inferred distance to the \ac{cGA} convergence point in \Manti; mass within the \ac{cGA} basin contributes ${\sim}75\%$ of the observer velocity amplitude, but the direction at this radius differs from the final direction by ${\sim}20\,\deg$, requiring contributions from larger scales to rotate the velocity vector.
    All error bars and shaded regions show $1\sigma$ uncertainties.
    The ``\texttt{Manticore} (obs)'' and ``\texttt{Manticore} (obs + $\sigma_v$)'' markers are offset horizontally for visual clarity.
    }
    \label{fig:manticore_velocity_sourcing}
\end{figure*}

In~\cref{fig:manticore_velocity_sourcing}, we examine which scales contribute to the \Manti observer velocity by applying the linear theory expression of~\cref{eq:linear_velocity} to compute the velocity sourced by matter within radius $R$.
Consistent with~\citet{Lavaux_2010}, mass within $40~\Mpch$ contributes approximately half the \ac{LG} velocity amplitude, broadly in line with the decomposition of~\citet{Tully_2008}, who attributed $259~\kmsec$ to evacuation from the Local Void, $185~\kmsec$ to the Virgo cluster and its surroundings, and $455~\kmsec$ toward Centaurus from scales beyond $3000~\kmsec$.
Even when including mass out to $155~\Mpch$ (the approximate extent of the \Manti reconstruction volume), we recover only $72 \pm 9\%$ of the \ac{CMB} dipole amplitude, with the direction remaining offset by $38 \pm 10\,\deg$.
Computing the velocity sourced specifically by mass within the \ac{cGA} basin (which is highly asymmetric around the observer; see~\cref{fig:ga_depth_sky}), we find $|\bm{V}_{\rm cGA}| = 324 \pm 51~\kmsec$ towards $(\ell,\,b) = (243 \pm 20,\,57 \pm 11)\,\deg$. This corresponds to ${\sim}75\%$ of the total observer velocity amplitude, but the direction differs by ${\sim}20\,\deg$ from the final velocity; mass beyond the basin is therefore required not only to increase the amplitude but also to rotate the velocity vector. Structures beyond the \ac{cGA} basin contribute significantly, with potentially substantial contributions from mass beyond the \Manti volume.

Two explanations may account for this discrepancy.
First, the measured \ac{CMB} dipole arises from the non-linear velocity field, whereas \Manti captures only mildly non-linear scales.
Following~\citet{Stiskalek_2025_VFO}, the residual velocity scatter between \ac{BORG} reconstructions and peculiar velocity data is typically ${\sim}150~\kmsec$; incorporating this scatter reduces the discrepancy to an insignificant $1\sigma$ disagreement, as shown in~\cref{fig:manticore_velocity_sourcing}.
However, this argument is disfavoured by the observation that all galaxies within the Local Sheet (${\lesssim}5~\Mpch$) share essentially the same velocity as the \ac{LG}~\citep{Tully_2008}, implying that the local flow field is dynamically cold and that the \ac{LG} velocity has no significant non-linear component.
Second, mass beyond ${\sim}150~\Mpch$ (the \Manti reconstruction volume), or in the \ac{ZoA}, may contribute significantly to the \ac{LG} velocity~\citep{Peacock_1992}.

\begin{figure}
    \centering
    \includegraphics[width=\columnwidth]{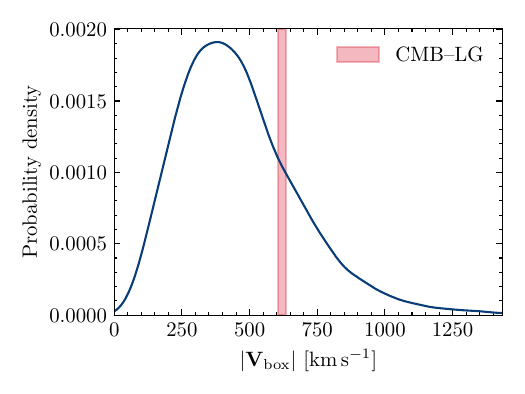}
    \caption{Distribution of observer velocity amplitudes $|\bm{V}_{\rm box}|$ across the $15{,}000$ fiducial-cosmology \texttt{Quijote} simulations~\protect\citep{Quijote}. Velocities are computed assuming linear theory (\cref{eq:linear_velocity}) for observers located at the box centre. The vertical red band indicates the observed \ac{LG} velocity in the \ac{CMB} frame of $620 \pm 15~\kmsec$ ($1\sigma$). The \ac{LG} velocity is consistent with typical \ac{LCDM} expectations.}
    \label{fig:quijote_vmag_histogram}
\end{figure}

\begin{figure*}
    \centering
    \includegraphics[width=\textwidth]{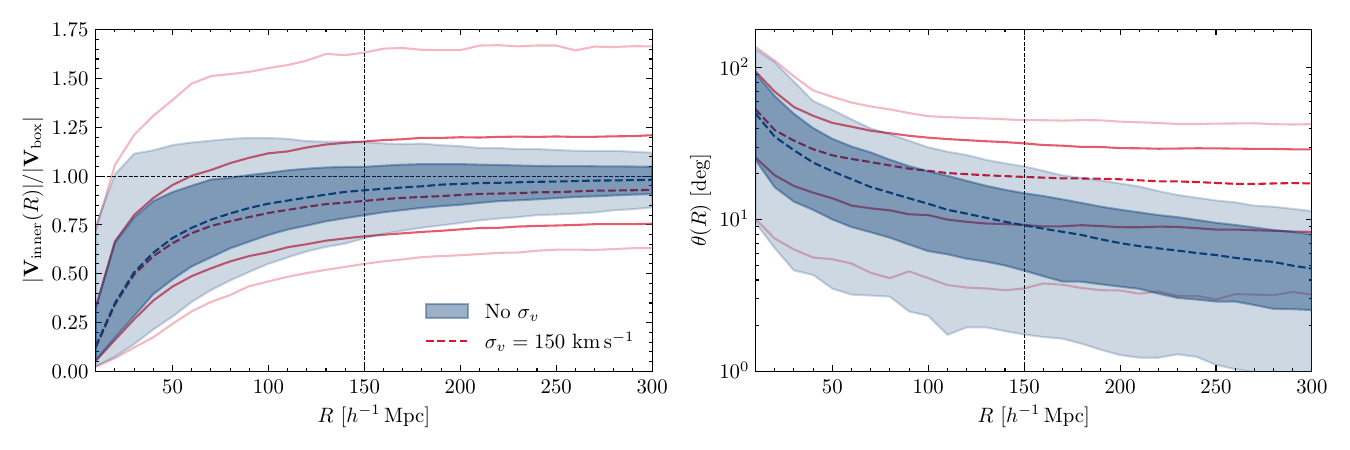}
    \caption{
    Convergence of observer velocity with reconstruction radius in random \ac{LCDM} simulations.
    We use these simulations to test whether the incomplete convergence of \Manti observer velocities at $155~\Mpch$ is consistent with \ac{LCDM} expectations for the contribution of mass beyond the reconstruction volume.
    Statistics are computed for observers with $|\bm{V}_{\rm box}|$ between $560$ and $680~\kmsec$, corresponding to the observed \ac{LG} velocity in the \ac{CMB} frame of $620 \pm 15~\kmsec$ at $4\sigma$.
    \textit{Left}: Normalised dot product $\bm{V}_{\rm inner}(R) \cdot \bm{V}_{\rm box} / (|\bm{V}_{\rm inner}(R)| |\bm{V}_{\rm box}|)$.
    \textit{Centre}: Amplitude ratio $|\bm{V}_{\rm inner}(R)| / |\bm{V}_{\rm box}|$.
    \textit{Right}: Alignment angle between $\bm{V}_{\rm inner}(R)$ and $\bm{V}_{\rm box}$.
    Bands show $1\sigma$ and $2\sigma$ confidence regions, with blue for pure linear velocities and red including $150~\kmsec$ scatter propagated to $\bm{V}_{\rm box}$ to mimic non-linear contributions.
    At $R = 150~\Mpch$, with scatter included, typical observers recover ${\sim}87\%$ of their total velocity amplitude with ${\sim}19\,\deg$ misalignment, demonstrating that the \Manti results are consistent with \ac{LCDM} expectations.
    }
    \label{fig:random_velocity_sourcing}
\end{figure*}

To test the second hypothesis, we analyse the $15{,}000$ random \texttt{Quijote} simulations described in Section~\ref{sec:random_sims}.
For each simulation, we extract the linear velocity of an observer at the box centre, $\bm{V}_{\rm box}$, and compute the velocity sourced by matter within radius $R$, $\bm{V}_{\rm inner}(R)$.
\Cref{fig:quijote_vmag_histogram} shows the distribution of $|\bm{V}_{\rm box}|$ across all simulations; the observed \ac{LG} velocity lies well within the bulk of the distribution, indicating that such velocities are typical in \ac{LCDM}.
In~\cref{fig:random_velocity_sourcing}, we select observers with $|\bm{V}_{\rm box}|$ between $560$ and $680~\kmsec$, corresponding to the observed \ac{LG} velocity in the \ac{CMB} frame of $620 \pm 15~\kmsec$~\citep{Planck_2018} at $4\sigma$; this yields $1{,}882$ observers out of $15{,}000$ ($12.5\%$).
We show the amplitude ratio $|\bm{V}_{\rm inner}(R)| / |\bm{V}_{\rm box}|$ and the angle between the two vectors (the normalised dot product $\bm{V}_{\rm inner}(R) \cdot \bm{V}_{\rm box} / |\bm{V}_{\rm box}|^2$ follows a nearly identical trend to the amplitude ratio and is omitted for brevity).
We present results for two cases: pure linear velocities, and velocities with $150~\kmsec$ noise added to $\bm{V}_{\rm box}$ to mimic non-linear contributions to the \ac{CMB} dipole.

\begin{figure}
    \centering
    \includegraphics[width=\columnwidth]{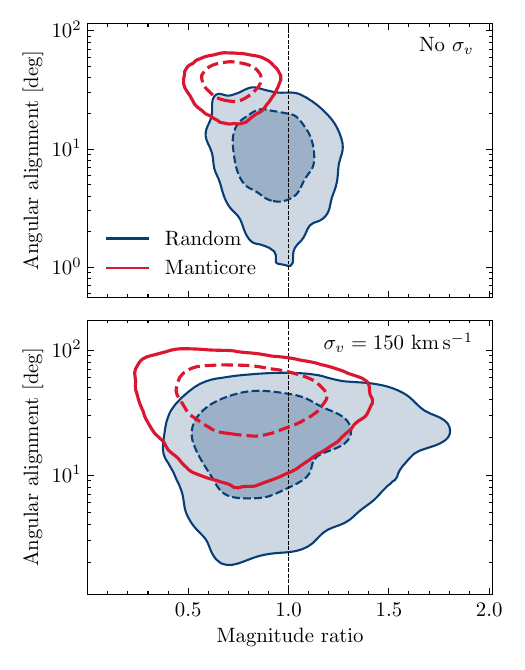}
    \caption{Joint distribution of magnitude ratio (velocity amplitude sourced within $155~\Mpch$ divided by reference velocity) and angular misalignment for \Manti (red contours) and random \texttt{Quijote} simulations (blue filled contours).
    For \Manti, the reference velocity is the observed \ac{CMB} dipole; for \texttt{Quijote}, it is the full-box velocity.
    Contours enclose $68\%$ and $95\%$ of the probability mass.
    \textit{Top}: Without small-scale velocity scatter $\sigma_v$, where \Manti is discrepant with random expectations at $2.0\sigma$ significance.
    \textit{Bottom}: Including $\sigma_v = 150~\kmsec$ scatter to account for non-linear contributions below the \ac{BORG} resolution, reducing the discrepancy to $0.5\sigma$.}
    \label{fig:velocity_convergence_scatter}
\end{figure}

For pure linear velocities at $R = 150~\Mpch$, we find $|\bm{V}_{\rm inner}| / |\bm{V}_{\rm box}| = 0.93^{+0.12}_{-0.13}$ and an alignment angle of $\theta = 9.2^{+5.8}_{-4.6}\,\deg$.
Although most of the observer velocity is typically sourced within this radius, notable outliers exist and misalignment can reach ${\sim}10\,\deg$.
Crucially, upon adding $150~\kmsec$ uncertainty to $\bm{V}_{\rm box}$, we find at $R = 150~\Mpch$ that $|\bm{V}_{\rm inner}| / |\bm{V}_{\rm box}| = 0.87^{+0.30}_{-0.18}$ and an alignment angle of $\theta = 19.1^{+12.7}_{-9.9}\,\deg$.
This scenario better represents actual observations, where the measured \ac{CMB} dipole contains non-linear contributions whilst the matter dipole is reconstructed only to mildly non-linear scales.
The \Manti results are therefore consistent with the observed \ac{LG} velocity in the \ac{CMB} frame at the ${\sim}1\sigma$ level.

We quantify this agreement more directly in~\cref{fig:velocity_convergence_scatter}, which shows the joint distribution of magnitude ratio and angular misalignment for both \Manti and the random \texttt{Quijote} simulations.
The magnitude ratio is defined as the amplitude of the velocity sourced by matter within $155~\Mpch$ divided by a reference velocity, while the angular misalignment is the angle between these two vectors. The reference velocity is taken to be the \ac{CMB} dipole for \Manti and the full-box velocity for \texttt{Quijote}; a direct comparison therefore assumes that the full velocity of a central observer in the cosmic rest frame is generated entirely within the \texttt{Quijote} volume.
Without small-scale velocity scatter $\sigma_v$, \Manti is discrepant with the random simulations at $2.0\sigma$ significance, computed using the posterior agreement metric.\footnote{\url{https://github.com/SebastianBocquet/PosteriorAgreement}}
This suggests that even including mass beyond the \Manti reconstruction volume is unlikely to fully reconcile the predicted and observed \ac{LG} velocities from linear theory alone.
However, upon adding $\sigma_v = 150~\kmsec$ scatter (representing small-scale, non-linear contributions to the observer velocity below the \ac{BORG} resolution), the discrepancy reduces to $0.5\sigma$.

\section{Discussion \& Conclusion}\label{sec:conclusion}

In this paper we examine the \ac{GA} using three complementary approaches: present-day streamline reconstruction (the traditional method yielding the \ac{cGA}), future gravitational evolution with \Manti digital twins built with the \ac{BORG} algorithm yielding the \ac{dGA}, and an assessment within \Manti of which scales source the \ac{LG} velocity in the \ac{CMB} frame.
\citet{Dressler_1987} fitted a bulk flow from elliptical-galaxy peculiar velocities and noted that Hydra--Centaurus itself was moving toward mass beyond ${\sim}5000~\kmsec$;~\citet{Bell_1988} then fitted an attractor model, identifying the \ac{GA} at $(\ell,\,b) \approx (307,\,9)\,\deg$ and $cz \approx 4350~\kmsec$ with inferred mass ${\sim}5.4 \times 10^{16}~\Msun$.
Recent peculiar-velocity samples~\citep{Tully_2014,Dupuy_2023} motivated streamline-based basin definitions~\citep{Dupuy_2019,Dupuy_2020}, with~\citet{Dupuy_2025} locating the \ac{cGA} at $(\ell,\,b) = (308.4 \pm 2.4\,\deg,\,29.0 \pm 1.9\,\deg)$ and $50 \pm 4~\Mpch$ via a joint machine-learning analysis combining redshift-space counts and peculiar velocities.

Applying the streamline method to \Manti at $\sigma_{\rm smooth} = 3~\Mpch$, we locate the \ac{cGA} convergence point at $(\ell,\,b) = (313^{+1}_{-5},\,27^{+2}_{-4})\,\deg$ and $41^{+2}_{-5}\,\Mpch$, close to Abell~3565 within the Hydra--Centaurus supercluster. The mass of the associated basin is $\log (M/(\Msunh))\!=\!16.4^{+0.1}_{-0.1}$ (\cref{tab:ga_positions}).
Hydra, Virgo, and Abell~3574 are \ac{cGA} basin members in most \Manti realisations, and the basin depth is asymmetric (\cref{fig:ga_depth_sky}).
Crucially,~\cref{fig:mw_streamlines_smoothing} shows that the streamline-defined convergence point shifts with smoothing: negligible smoothing yields Virgo, intermediate smoothing the \ac{cGA}, and large smoothing Shapley, underscoring the construct’s ambiguity.
Abell~3565 and Abell~S0753 counterparts typically merge by $a=10$, while Centaurus remains near its present position.
We summarise the \Manti counterparts to these observed clusters in \cref{tab:manticore_clusters}.
Norma, though massive, is not part of the \ac{cGA} but rather generally forms its own basin.

Evolving the \Manti realisations to $a=10$ from the present in the \ac{BPT} suite (corresponding to a Universe age of $52.4~{\rm Gyr}$), particles within $5~\cMpch$ of the observer drift only $3 \pm 1~\cMpch$ toward Virgo, remaining unbound; simulating to the scale factor of 100 changes nothing because accelerated expansion halts further structure growth (\cref{fig:observer_particle_displacements}).
Virgo itself shifts modestly to $r \approx 14~\cMpch$ and $(\ell,\,b) \approx (284,\,67)\,\deg$ by $a=10$.
We find that Virgo dominates the gravitational pull on the observer, defining the future \emph{dynamical GA} direction, with Perseus, Fornax, Ursa Major, and Norma contributing at roughly one-fifth of Virgo's level, and Centaurus at ${\sim}13\%$ (\cref{tab:force_contributions}).
However, particles from the observer position never actually reach Virgo; they travel only ${\sim}3~\cMpch$ whilst Virgo remains ${\sim}14~\cMpch$ away at $a=10$, with dark energy domination within the assumed flat \ac{LCDM} cosmology preventing further gravitational collapse.

Addressing the original motivation of~\citet{Bell_1988}---whether the \ac{cGA} sources the \ac{LG} peculiar velocity---we find that mass within $155~\Mpch$ (the extent of the \Manti reconstruction volume) accounts for only $72 \pm 9\%$ of the \ac{CMB} dipole amplitude, with the direction offset by $38 \pm 10\,\deg$.
Using random \ac{LCDM} simulations, we show that this incomplete convergence is consistent with expectations: considering the velocity sourced by matter within radius $R$, observers with \ac{LG}-like velocities receive $93 \pm 12\%$ of their total velocity amplitude with $9.2 \pm 5\,\deg$ misalignment from sources within $R = 150~\Mpch$ for pure linear velocities. Upon adding $150~\kmsec$ scatter to mimic non-linear contributions to the \ac{CMB} dipole, the amplitude ratio becomes $87^{+30}_{-18}\%$ with $19.1^{+12.7}_{-9.9}\,\deg$ misalignment.
Directly comparing \Manti to the random simulations in the joint space of magnitude ratio and angular misalignment, the discrepancy is $2.0\sigma$ without small-scale velocity scatter $\sigma_v$, reducing to $0.5\sigma$ upon adding $\sigma_v = 150~\kmsec$.

It is therefore not possible to identify the origin of the \ac{LG} velocity from linear or mildly non-linear reconstructions of the local Universe, and nor is it expected to be~\citep{Peacock_1992}.
This is due to (i) mass beyond the \Manti reconstruction volume, which deeper \ac{BORG} reconstructions extending beyond the \TWOMPP could constrain, and (ii) small-scale non-linear motions below the \ac{BORG} resolution, since comparisons between linear and non-linear velocity fields typically yield ${\sim}150~\kmsec$ scatter~\citep{Carrick_2015}, and the \ac{LG} itself (with total mass ${\sim}3 \times 10^{12}~\Msun$;~\citealt{Benisty_2022_LGmass,Wempe_2024}) lies well below the \ac{BORG} constraint scale.
Higher-resolution simulations of the immediate \ac{LG} neighbourhood, such as the \ac{BORG}-based reconstructions of~\citet{Wempe_2024,Wempe_2025}, could help to quantify the non-linear contribution.

In summary, a \ac{cGA} convergence point appears in present-day streamlines but only at intermediate smoothing.
Furthermore, the \ac{cGA} does not determine the \ac{LG}'s future displacement direction; among bound structures, Virgo dominates, with other clusters contributing at only $10$--$20\%$ of its influence, and the \ac{LG} never reaches Virgo unless physics beyond \ac{LCDM} alters the expansion history.
Addressing the original question of what structures source the ${\sim}600~\kmsec$ \ac{CMB} dipole, mass within $155~\Mpch$ recovers only ${\sim}72\%$ of the amplitude with ${\sim}38\,\deg$ directional offset; the Virgo supercluster alone contributes only ${\sim}200~\kmsec$, with the remainder arising from integration over larger scales (Section~\ref{sec:lg_velocity}).
Mass within the asymmetric \ac{cGA} basin contributes ${\sim}75\%$ of the observer's linear velocity amplitude, though the direction at this radius is offset by ${\sim}20\,\deg$ from the final velocity, with matter beyond the basin required to rotate the velocity vector.

These conclusions rest on the \Manti realisations, the adopted smoothing and resolution, and the assumption of a flat \ac{LCDM} cosmology.
It is remarkable that \Manti realisations provide counterpart haloes for nearly all observed clusters in the \ac{GA} basin, including nearby lower-mass systems such as Fornax and Ursa Major, demonstrating the power of \ac{BORG}-enabled digital twins for high-fidelity studies of the local Universe.
These simulations accurately reproduce the \ac{cGA} convergence point and allow us to pinpoint its location and the cluster population of its basin, showing that although it is a substantial overdensity it is not as overdense as the region surrounding the Norma and Perseus clusters.
They also reveal that Norma is not a member of the streamline basin and that the \ac{cGA} does not dominate the future dynamics of the \ac{LG}.
The streamline convergence point is thus a scale-dependent property of the instantaneous velocity field rather than a unique physical structure, and does not explain the local dynamics.
With perfect knowledge of the velocity field, a unique set of basins of attraction, each with its own sink, would in principle be well defined. In practice, however, any finite-resolution reconstruction smooths over small-scale structure, merging smaller basins and their sinks into larger ones and making smoothing-scale-dependence unavoidable.
Moreover, when dark energy becomes dominant and perturbation growth freezes, present-day streamlines become particularly poor tracers of future trajectories.

We therefore conclude that: (i) Virgo dominates the \acl{LG}'s immediate dynamical evolution; (ii) the \acl{cGA} represents a moderate overdensity embedded within a multiscale flow; and (iii) addressing the original motivation of the \ac{GA}---explaining the \ac{CMB} dipole---the \ac{LG} velocity is sourced by structures across a wide range of scales, with even the linear contribution not guaranteed to converge within the ${\sim}150~\Mpch$ volume probed here, and the remaining discrepancy likely arising from mass beyond this volume and small-scale non-linear motions.
The \ac{BORG} methodology of inferring initial conditions via Bayesian field-level reconstruction and evolving them beyond the present opens new avenues for understanding cosmic dynamics and could be extended to joint analyses with peculiar-velocity data to further constrain structures behind the \ac{ZoA}, and more generally to understand both the density and velocity fields of the local Universe.

\section*{Acknowledgments}

We thank Pedro Ferreira for useful inputs and discussions.

\vspace{1em}

RS acknowledges financial support from STFC Grant No. ST/X508664/1 and the Snell Exhibition of Balliol College, Oxford.
HD is supported by a Royal Society University Research Fellowship (grant no. 211046).
JJ acknowledges support from the research project \emph{Understanding the Dynamic Universe}, funded by the Knut and Alice Wallenberg Foundation (Dnr KAW~2018.0067).
JJ acknowledges the hospitality of the Aspen Center for Physics, which is supported by National Science Foundation grant PHY-1607611, and support for participation at the Aspen Center provided by the Simons Foundation.
JJ acknowledges support from the Swedish Research Council (VR) under project 2020-05143, \emph{Deciphering the Dynamics of Cosmic Structure}.
GL, JJ, and SM acknowledge support from the Simons Foundation through the Simons Collaboration on ``Learning the Universe''.
GL acknowledges support from the CNRS IEA programme ``Manticore''.
MJH acknowledges support from an NSERC Discovery grant.

This work was done within the Aquila Consortium\footnote{\url{https://aquila-consortium.org}}.
The authors would like to acknowledge the use of the University of Oxford Advanced Research Computing (ARC) facility in carrying out this work\footnote{\url{https://doi.org/10.5281/zenodo.22558}}.
This research has made use of the NASA/IPAC Extragalactic Database (NED), which is operated by the Jet Propulsion Laboratory, California Institute of Technology, under contract with the National Aeronautics and Space Administration.

\bibliographystyle{apsrev4-1}
\bibliography{ref}

@ARTICLE{Tully_2008,
       author = {{Tully}, R. Brent and {Shaya}, Edward J. and {Karachentsev}, Igor D. and {Courtois}, H{\'e}l{\`e}ne M. and {Kocevski}, Dale D. and {Rizzi}, Luca and {Peel}, Alan},
        title = "{Our Peculiar Motion Away from the Local Void}",
      journal = {\apj},
         year = 2008,
        month = mar,
       volume = {676},
       number = {1},
        pages = {184-205},
          doi = {10.1086/527428},
archivePrefix = {arXiv},
       eprint = {0705.4139},
}

@ARTICLE{Pomarede_2020,
       author = {{Pomar{\`e}de}, Daniel and {Tully}, R. Brent and {Graziani}, Romain and {Courtois}, H{\'e}l{\`e}ne M. and {Hoffman}, Yehuda and {Lezmy}, J{\'e}r{\'e}my},
        title = "{Cosmicflows-3: The South Pole Wall}",
      journal = {\apj},
         year = 2020,
        month = jul,
       volume = {897},
       number = {2},
          eid = {133},
        pages = {133},
          doi = {10.3847/1538-4357/ab9952},
archivePrefix = {arXiv},
       eprint = {2007.04414},
 primaryClass = {astro-ph.CO}
}

@ARTICLE{Pomarede_2017,
       author = {{Pomar{\`e}de}, Daniel and {Hoffman}, Yehuda and {Courtois}, H{\'e}l{\`e}ne M. and {Tully}, R. Brent},
        title = "{The Cosmic V-Web}",
      journal = {\apj},
         year = 2017,
        month = aug,
       volume = {845},
       number = {1},
          eid = {55},
        pages = {55},
          doi = {10.3847/1538-4357/aa7f78},
archivePrefix = {arXiv},
       eprint = {1706.03413},
 primaryClass = {astro-ph.CO}
}

@ARTICLE{Hollinger_2024,
       author = {{Hollinger}, Amber M. and {Hudson}, Michael J.},
        title = "{Cosmological parameters estimated from peculiar velocity-density comparisons: calibrating 2M++}",
      journal = {\mnras},
         year = 2024,
        month = jun,
       volume = {531},
       number = {1},
        pages = {788-804},
          doi = {10.1093/mnras/stae1042},
archivePrefix = {arXiv},
       eprint = {2312.03904},
 primaryClass = {astro-ph.CO}
}

@ARTICLE{Shaya_1995,
       author = {{Shaya}, Edward J. and {Peebles}, P.~J.~E. and {Tully}, R. Brent},
        title = "{Action Principle Solutions for Galaxy Motions within 3000 Kilometers per Second}",
      journal = {\apj},
         year = 1995,
        month = nov,
       volume = {454},
        pages = {15},
          doi = {10.1086/176460},
archivePrefix = {arXiv},
       eprint = {astro-ph/9506144},
 primaryClass = {astro-ph}
}

@ARTICLE{Shaya_2017,
       author = {{Shaya}, Edward J. and {Tully}, R. Brent and {Hoffman}, Yehuda and {Pomar{\`e}de}, Daniel},
        title = "{Action Dynamics of the Local Supercluster}",
      journal = {\apj},
         year = 2017,
        month = dec,
       volume = {850},
       number = {2},
          eid = {207},
        pages = {207},
          doi = {10.3847/1538-4357/aa9525},
archivePrefix = {arXiv},
       eprint = {1710.08935},
 primaryClass = {astro-ph.CO}
}

@ARTICLE{Shaya_2022,
       author = {{Shaya}, Edward J. and {Tully}, R. Brent and {Pomarede}, Daniel and {Peel}, Alan},
        title = "{Galaxy Flows within 8000 km s$^{-1}$ from Numerical Action Methods}",
      journal = {\apj},
         year = 2022,
        month = mar,
       volume = {927},
       number = {2},
          eid = {168},
        pages = {168},
          doi = {10.3847/1538-4357/ac4f66},
archivePrefix = {arXiv},
       eprint = {2201.12315},
 primaryClass = {astro-ph.CO}
}

@ARTICLE{McAlpine2025_ManticoreCatalogue,
       author = {{McAlpine}, Stuart},
        title = "{The Manticore-Local Cluster Catalogue: A Posterior Map of Massive Structures in the Nearby Universe}",
      journal = {arXiv e-prints},
         year = 2025,
        month = oct,
          eid = {arXiv:2510.16574},
        pages = {arXiv:2510.16574},
          doi = {10.48550/arXiv.2510.16574},
archivePrefix = {arXiv},
       eprint = {2510.16574},
 primaryClass = {astro-ph.CO},
       adsurl = {https://ui.adsabs.harvard.edu/abs/2025arXiv251016574M}
}

@ARTICLE{Planck_2020_cosmo,
       author = {{Planck Collaboration} and {Aghanim}, N. and {Akrami}, Y. and {Ashdown}, M. and {Aumont}, J. and {Baccigalupi}, C. and {Ballardini}, M. and {Banday}, A.~J. and {Barreiro}, R.~B. and {Bartolo}, N. and others},
        title = "{Planck 2018 results. VI. Cosmological parameters}",
      journal = {\aap},
         year = 2020,
        month = sep,
       volume = {641},
          eid = {A6},
        pages = {A6},
          doi = {10.1051/0004-6361/201833910},
archivePrefix = {arXiv},
       eprint = {1807.06209},
 primaryClass = {astro-ph.CO},
       adsurl = {https://ui.adsabs.harvard.edu/abs/2020A&A...641A...6P}
}

@ARTICLE{Jasche_2019,
       author = {{Jasche}, J. and {Lavaux}, G.},
        title = "{Physical Bayesian modelling of the non-linear matter distribution: New insights into the nearby universe}",
      journal = {\aap},
         year = 2019,
        month = may,
       volume = {625},
          eid = {A64},
        pages = {A64},
          doi = {10.1051/0004-6361/201833710},
archivePrefix = {arXiv},
       eprint = {1806.11117},
 primaryClass = {astro-ph.CO},
       adsurl = {https://ui.adsabs.harvard.edu/abs/2019A&A...625A..64J}
}

@ARTICLE{Stopyra_2023,
       author = {{Stopyra}, Stephen and {Peiris}, Hiranya V. and {Pontzen}, Andrew and {Jasche}, Jens and {Lavaux}, Guilhem},
        title = "{Towards accurate field-level inference of massive cosmic structures}",
      journal = {\mnras},
         year = 2024,
        month = jan,
       volume = {527},
       number = {1},
        pages = {1244-1256},
          doi = {10.1093/mnras/stad3170},
archivePrefix = {arXiv},
       eprint = {2304.09193},
 primaryClass = {astro-ph.CO},
       adsurl = {https://ui.adsabs.harvard.edu/abs/2024MNRAS.527.1244S}
}

@ARTICLE{Jasche_2013,
       author = {{Jasche}, Jens and {Wandelt}, Benjamin D.},
        title = "{Bayesian physical reconstruction of initial conditions from large-scale structure surveys}",
      journal = {\mnras},
         year = 2013,
        month = jun,
       volume = {432},
       number = {2},
        pages = {894-913},
          doi = {10.1093/mnras/stt449},
archivePrefix = {arXiv},
       eprint = {1203.3639},
 primaryClass = {astro-ph.CO},
       adsurl = {https://ui.adsabs.harvard.edu/abs/2013MNRAS.432..894J}
}

@ARTICLE{Lavaux_2011,
       author = {{Lavaux}, Guilhem and {Hudson}, Michael J.},
        title = "{The 2M++ galaxy redshift catalogue}",
      journal = {\mnras},
         year = 2011,
        month = oct,
       volume = {416},
       number = {4},
        pages = {2840-2856},
          doi = {10.1111/j.1365-2966.2011.19233.x},
archivePrefix = {arXiv},
       eprint = {1105.6107},
 primaryClass = {astro-ph.CO},
       adsurl = {https://ui.adsabs.harvard.edu/abs/2011MNRAS.416.2840L}
}

@ARTICLE{Lavaux_2016,
       author = {{Lavaux}, Guilhem and {Jasche}, Jens},
        title = "{Unmasking the masked Universe: the 2M++ catalogue through Bayesian eyes}",
      journal = {\mnras},
         year = 2016,
        month = jan,
       volume = {455},
       number = {3},
        pages = {3169-3179},
          doi = {10.1093/mnras/stv2499},
archivePrefix = {arXiv},
       eprint = {1509.05040},
 primaryClass = {astro-ph.CO},
       adsurl = {https://ui.adsabs.harvard.edu/abs/2016MNRAS.455.3169L}
}

@ARTICLE{Lavaux_2019,
       author = {{Lavaux}, Guilhem and {Jasche}, Jens and {Leclercq}, Florent},
        title = "{Systematic-free inference of the cosmic matter density field from SDSS3-BOSS data}",
      journal = {arXiv e-prints},
         year = 2019,
        month = sep,
          eid = {arXiv:1909.06396},
        pages = {arXiv:1909.06396},
          doi = {10.48550/arXiv.1909.06396},
archivePrefix = {arXiv},
       eprint = {1909.06396},
 primaryClass = {astro-ph.CO},
       adsurl = {https://ui.adsabs.harvard.edu/abs/2019arXiv190906396L}
}

@ARTICLE{Porqueres_2019,
       author = {{Porqueres}, Natalia and {Kodi Ramanah}, Doogesh and {Jasche}, Jens and {Lavaux}, Guilhem},
        title = "{Explicit Bayesian treatment of unknown foreground contaminations in galaxy surveys}",
      journal = {\aap},
         year = 2019,
        month = apr,
       volume = {624},
          eid = {A115},
        pages = {A115},
          doi = {10.1051/0004-6361/201834844},
archivePrefix = {arXiv},
       eprint = {1812.05113},
 primaryClass = {astro-ph.CO},
       adsurl = {https://ui.adsabs.harvard.edu/abs/2019A&A...624A.115P}
}

@ARTICLE{Planck_2020,
       author = {{Planck Collaboration} and {Aghanim}, N. and {Akrami}, Y. and others},
        title = "{Planck 2018 results. I. Overview and the cosmological legacy of Planck}",
      journal = {\aap},
         year = 2020,
        month = sep,
       volume = {641},
          eid = {A1},
        pages = {A1},
          doi = {10.1051/0004-6361/201833880},
archivePrefix = {arXiv},
       eprint = {1807.06205},
 primaryClass = {astro-ph.CO},
       adsurl = {https://ui.adsabs.harvard.edu/abs/2020A&A...641A...1P}
}

@ARTICLE{Stiskalek_2025_VFO,
       author = {{Stiskalek}, Richard and {Desmond}, Harry and {Devriendt}, Julien and {Slyz}, Adrianne and {Lavaux}, Guilhem and {Hudson}, Michael J. and {Bartlett}, Deaglan J. and {Courtois}, H{\'e}l{\`e}ne M.},
        title = "{The Velocity Field Olympics: Assessing velocity field reconstructions with direct distance tracers}",
      journal = {\mnras},
         year = 2025,
        month = nov,
          doi = {10.1093/mnras/staf1960},
archivePrefix = {arXiv},
       eprint = {2502.00121},
 primaryClass = {astro-ph.CO},
       adsurl = {https://ui.adsabs.harvard.edu/abs/2025MNRAS.tmp.1852S}
}

@ARTICLE{Gadget4,
       author = {{Springel}, Volker and {Pakmor}, R{\"u}diger and {Zier}, Oliver and {Reinecke}, Martin},
        title = "{Simulating cosmic structure formation with the GADGET-4 code}",
      journal = {\mnras},
         year = 2021,
        month = sep,
       volume = {506},
       number = {2},
        pages = {2871-2949},
          doi = {10.1093/mnras/stab1855},
archivePrefix = {arXiv},
       eprint = {2010.03567},
 primaryClass = {astro-ph.IM},
       adsurl = {https://ui.adsabs.harvard.edu/abs/2021MNRAS.506.2871S}
}

@ARTICLE{DES_Y3,
       author = {{Abbott}, T.~M.~C. and {Aguena}, M. and {Alarcon}, A. and {Allam}, S. and {Alves}, O. and {Amon}, A. and {Andrade-Oliveira}, F. and {Annis}, J. and {Avila}, S. and {Bacon}, D. and others},
        title = "{Dark Energy Survey Year 3 results: Cosmological constraints from galaxy clustering and weak lensing}",
      journal = {\prd},
         year = 2022,
        month = jan,
       volume = {105},
       number = {2},
          eid = {023520},
        pages = {023520},
          doi = {10.1103/PhysRevD.105.023520},
archivePrefix = {arXiv},
       eprint = {2105.13549},
 primaryClass = {astro-ph.CO},
       adsurl = {https://ui.adsabs.harvard.edu/abs/2022PhRvD.105b3520A}
}

@ARTICLE{Schaller_2024,
       author = {{Schaller}, Matthieu and {Borrow}, Josh and {Draper}, Peter W. and {Ivkovic}, Mladen and {McAlpine}, Stuart and {Vandenbroucke}, Bert and {Bah{\'e}}, Yannick and {Chaikin}, Evgenii and {Chalk}, Aidan B.~G. and {Chan}, Tsang Keung and {Correa}, Camila and {van Daalen}, Marcel and {Elbers}, Willem and {Gonnet}, Pedro and {Hausammann}, Lo{\"\i}c and {Helly}, John and {Hu{\v{s}}ko}, Filip and {Kegerreis}, Jacob A. and {Nobels}, Folkert S.~J. and {Ploeckinger}, Sylvia and {Revaz}, Yves and {Roper}, William J. and {Ruiz-Bonilla}, Sergio and {Sandnes}, Thomas D. and {Uyttenhove}, Yolan and {Willis}, James S. and {Xiang}, Zhen},
        title = "{SWIFT: A modern highly-parallel gravity and smoothed particle hydrodynamics solver for astrophysical and cosmological applications}",
      journal = {\mnras},
         year = 2024,
        month = may,
       volume = {530},
       number = {2},
        pages = {2378-2419},
          doi = {10.1093/mnras/stae922},
archivePrefix = {arXiv},
       eprint = {2305.13380},
 primaryClass = {astro-ph.IM},
       adsurl = {https://ui.adsabs.harvard.edu/abs/2024MNRAS.530.2378S}
}

@ARTICLE{Jasche_2015,
       author = {{Jasche}, J. and {Leclercq}, F. and {Wandelt}, B.~D.},
        title = "{Past and present cosmic structure in the SDSS DR7 main sample}",
      journal = {\jcap},
         year = 2015,
        month = jan,
       volume = {2015},
       number = {1},
        pages = {036-036},
          doi = {10.1088/1475-7516/2015/01/036},
archivePrefix = {arXiv},
       eprint = {1409.6308},
 primaryClass = {astro-ph.CO},
       adsurl = {https://ui.adsabs.harvard.edu/abs/2015JCAP...01..036J}
}

@ARTICLE{Leclercq_2017,
       author = {{Leclercq}, Florent and {Jasche}, Jens and {Lavaux}, Guilhem and {Wandelt}, Benjamin and {Percival}, Will},
        title = "{The phase-space structure of nearby dark matter as constrained by the SDSS}",
      journal = {\jcap},
         year = 2017,
        month = jun,
       volume = {2017},
       number = {6},
          eid = {049},
        pages = {049},
          doi = {10.1088/1475-7516/2017/06/049},
archivePrefix = {arXiv},
       eprint = {1601.00093},
 primaryClass = {astro-ph.CO},
       adsurl = {https://ui.adsabs.harvard.edu/abs/2017JCAP...06..049L}
}

@ARTICLE{McAlpine_2025,
       author = {{McAlpine}, Stuart and {Jasche}, Jens and {Ata}, Metin and {Lavaux}, Guilhem and {Stiskalek}, Richard and {Frenk}, Carlos S. and {Jenkins}, Adrian},
        title = "{The Manticore Project I: a digital twin of our cosmic neighbourhood from Bayesian field-level analysis}",
      journal = {\mnras},
         year = 2025,
        month = jun,
       volume = {540},
       number = {1},
        pages = {716-745},
          doi = {10.1093/mnras/staf767},
archivePrefix = {arXiv},
       eprint = {2505.10682},
 primaryClass = {astro-ph.CO},
       adsurl = {https://ui.adsabs.harvard.edu/abs/2025MNRAS.540..716M}
}

@ARTICLE{Dupuy_2025,
       author = {{Dupuy}, Alexandra and {Jeong}, Donghui and {Hong}, Sungwook E. and {Hwang}, Ho Seong and {Kim}, Juhan and {Courtois}, H{\'e}l{\`e}ne M.},
        title = "{Revealing Hidden Cosmic Flows through the Zone of Avoidance with Deep Learning}",
      journal = {arXiv e-prints},
         year = 2025,
        month = nov,
          eid = {arXiv:2511.03919},
        pages = {arXiv:2511.03919},
          doi = {10.48550/arXiv.2511.03919},
archivePrefix = {arXiv},
       eprint = {2511.03919},
 primaryClass = {astro-ph.CO},
       adsurl = {https://ui.adsabs.harvard.edu/abs/2025arXiv251103919D}
}

@ARTICLE{Dupuy_2020,
       author = {{Dupuy}, Alexandra and {Courtois}, H{\'e}l{\`e}ne M. and {Libeskind}, Noam I. and {Guinet}, Daniel},
        title = "{Segmenting the Universe into dynamically coherent basins}",
      journal = {\mnras},
         year = 2020,
        month = apr,
       volume = {493},
       number = {3},
        pages = {3513-3520},
          doi = {10.1093/mnras/staa536},
archivePrefix = {arXiv},
       eprint = {2002.06814},
 primaryClass = {astro-ph.CO},
       adsurl = {https://ui.adsabs.harvard.edu/abs/2020MNRAS.493.3513D}
}

@article{Smoot,
  title = {Detection of Anisotropy in the Cosmic Blackbody Radiation},
  author = {Smoot, G. F. and Gorenstein, M. V. and Muller, R. A.},
  journal = {Phys. Rev. Lett.},
  volume = {39},
  issue = {14},
  pages = {898--901},
  numpages = {0},
  year = {1977},
  month = {Oct},
  publisher = {American Physical Society},
  doi = {10.1103/PhysRevLett.39.898},
  url = {https://link.aps.org/doi/10.1103/PhysRevLett.39.898}
}

@ARTICLE{Dupuy_2019,
       author = {{Dupuy}, Alexandra and {Courtois}, Helene M. and {Dupont}, Florent and {Denis}, Florence and {Graziani}, Romain and {Copin}, Yannick and {Pomar{\`e}de}, Daniel and {Libeskind}, Noam and {Carlesi}, Edoardo and {Tully}, Brent and {Guinet}, Daniel},
        title = "{Partitioning the Universe into gravitational basins using the cosmic velocity field}",
      journal = {\mnras},
         year = 2019,
        month = oct,
       volume = {489},
       number = {1},
        pages = {L1-L6},
          doi = {10.1093/mnrasl/slz115},
archivePrefix = {arXiv},
       eprint = {1907.06555},
 primaryClass = {astro-ph.CO},
       adsurl = {https://ui.adsabs.harvard.edu/abs/2019MNRAS.489L...1D}
}

@ARTICLE{Valade_2024,
       author = {{Valade}, A. and {Libeskind}, N.~I. and {Pomar{\`e}de}, D. and {Tully}, R.~B. and {Hoffman}, Y. and {Pfeifer}, S. and {Kourkchi}, E.},
        title = "{Identification of basins of attraction in the local Universe}",
      journal = {Nature Astronomy},
         year = 2024,
        month = dec,
       volume = {8},
        pages = {1610-1616},
          doi = {10.1038/s41550-024-02370-0},
archivePrefix = {arXiv},
       eprint = {2409.17261},
 primaryClass = {astro-ph.CO},
       adsurl = {https://ui.adsabs.harvard.edu/abs/2024NatAs...8.1610V}
}

@ARTICLE{Dressler_1987,
       author = {{Dressler}, Alan and {Faber}, S.~M. and {Burstein}, David and {Davies}, Roger L. and {Lynden-Bell}, Donald and {Terlevich}, R.~J. and {Wegner}, Gary},
        title = "{Spectroscopy and Photometry of Elliptical Galaxies: A Large-Scale Streaming Motion in the Local Universe}",
      journal = {\apjl},
         year = 1987,
        month = feb,
       volume = {313},
        pages = {L37},
          doi = {10.1086/184827},
  adsurl = {https://ui.adsabs.harvard.edu/abs/1987ApJ...313L..37D}
}

@ARTICLE{Bell_1988,
       author = {{Lynden-Bell}, D. and {Faber}, S.~M. and {Burstein}, David and {Davies}, Roger L. and {Dressler}, Alan and {Terlevich}, R.~J. and {Wegner}, Gary},
        title = "{Photometry and Spectroscopy of Elliptical Galaxies. V. Galaxy Streaming toward the New Supergalactic Center}",
      journal = {\apj},
         year = 1988,
        month = mar,
       volume = {326},
        pages = {19},
          doi = {10.1086/166066},
       adsurl = {https://ui.adsabs.harvard.edu/abs/1988ApJ...326...19L}
}

@ARTICLE{Dupuy_2023,
       author = {{Dupuy}, A. and {Courtois}, H.~M.},
        title = "{Dynamic cosmography of the local Universe: Laniakea and five more watershed superclusters}",
      journal = {\aap},
         year = 2023,
        month = oct,
       volume = {678},
          eid = {A176},
        pages = {A176},
          doi = {10.1051/0004-6361/202346802},
archivePrefix = {arXiv},
       eprint = {2305.02339},
 primaryClass = {astro-ph.CO},
       adsurl = {https://ui.adsabs.harvard.edu/abs/2023A&A...678A.176D}
}

@ARTICLE{Davis_1985_FoF,
       author = {{Davis}, M. and {Efstathiou}, G. and {Frenk}, C.~S. and {White}, S.~D.~M.},
        title = "{The evolution of large-scale structure in a universe dominated by cold dark matter}",
      journal = {\apj},
         year = 1985,
        month = may,
       volume = {292},
        pages = {371-394},
          doi = {10.1086/163168},
       adsurl = {https://ui.adsabs.harvard.edu/abs/1985ApJ...292..371D}
}

@ARTICLE{Monaghan_1992,
       author = {{Monaghan}, J.~J.},
        title = "{Smoothed particle hydrodynamics.}",
      journal = {\araa},
         year = 1992,
        month = jan,
       volume = {30},
        pages = {543-574},
          doi = {10.1146/annurev.aa.30.090192.002551},
       adsurl = {https://ui.adsabs.harvard.edu/abs/1992ARA&A..30..543M}
}

@ARTICLE{Colombi_2007,
       author = {{Colombi}, St{\'e}phane and {Chodorowski}, Micha{\l} J. and {Teyssier}, Romain},
        title = "{Cosmic velocity-gravity relation in redshift space}",
      journal = {\mnras},
         year = 2007,
        month = feb,
       volume = {375},
       number = {1},
        pages = {348-370},
          doi = {10.1111/j.1365-2966.2006.11330.x},
archivePrefix = {arXiv},
       eprint = {0805.1693},
 primaryClass = {astro-ph},
       adsurl = {https://ui.adsabs.harvard.edu/abs/2007MNRAS.375..348C}
}

@ARTICLE{Planck_2018,
       author = {{Planck Collaboration}},
        title = "{Planck 2018 results. I. Overview and the cosmological legacy of Planck}",
      journal = {\aap},
         year = 2020,
        month = sep,
       volume = {641},
          eid = {A1},
        pages = {A1},
          doi = {10.1051/0004-6361/201833880},
archivePrefix = {arXiv},
       eprint = {1807.06205},
 primaryClass = {astro-ph.CO},
       adsurl = {https://ui.adsabs.harvard.edu/abs/2020A&A...641A...1P}
}

@INPROCEEDINGS{Woudt_2000,
       author = {{Woudt}, P.~A. and {Kraan-Korteweg}, R.~C.},
        title = "{Large-Scale Structures Behind the Southern Milky Way in the Great Attractor Region}",
    booktitle = {Mapping the Hidden Universe: The Universe behind the Mily Way - The Universe in HI},
         year = 2000,
       editor = {{Kraan-Korteweg}, Ren{\'e}e C. and {Henning}, Patricia A. and {Andernach}, Heinz},
       series = {Astronomical Society of the Pacific Conference Series},
       volume = {218},
        month = jan,
        pages = {193},
          doi = {10.48550/arXiv.astro-ph/0006126},
archivePrefix = {arXiv},
       eprint = {astro-ph/0006126},
 primaryClass = {astro-ph},
       adsurl = {https://ui.adsabs.harvard.edu/abs/2000ASPC..218..193W}
}

@ARTICLE{Woudt_2008,
       author = {{Woudt}, P.~A. and {Kraan-Korteweg}, R.~C. and {Lucey}, J. and {Fairall}, A.~P. and {Moore}, S.~A.~W.},
        title = "{The Norma cluster (ACO 3627) - I. A dynamical analysis of the most massive cluster in the Great Attractor}",
      journal = {\mnras},
         year = 2008,
        month = jan,
       volume = {383},
       number = {2},
        pages = {445-457},
          doi = {10.1111/j.1365-2966.2007.12571.x},
archivePrefix = {arXiv},
       eprint = {0706.2227},
 primaryClass = {astro-ph},
       adsurl = {https://ui.adsabs.harvard.edu/abs/2008MNRAS.383..445W}
}

@ARTICLE{Trentham_2001,
       author = {{Trentham}, Neil and {Tully}, R. Brent and {Verheijen}, Marc A.~W.},
        title = "{The Ursa Major cluster of galaxies - III. Optical observations of dwarf galaxies and the luminosity function down to M$_{R}$=-11}",
      journal = {\mnras},
         year = 2001,
        month = jul,
       volume = {325},
       number = {1},
        pages = {385-404},
          doi = {10.1046/j.1365-8711.2001.04427.x},
archivePrefix = {arXiv},
       eprint = {astro-ph/0103039},
 primaryClass = {astro-ph},
       adsurl = {https://ui.adsabs.harvard.edu/abs/2001MNRAS.325..385T}
}

@ARTICLE{Tully_2014,
       author = {{Tully}, R. Brent and {Courtois}, H{\'e}l{\`e}ne and {Hoffman}, Yehuda and {Pomar{\`e}de}, Daniel},
        title = "{The Laniakea supercluster of galaxies}",
      journal = {\nat},
         year = 2014,
        month = sep,
       volume = {513},
       number = {7516},
        pages = {71-73},
          doi = {10.1038/nature13674},
archivePrefix = {arXiv},
       eprint = {1409.0880},
 primaryClass = {astro-ph.CO},
       adsurl = {https://ui.adsabs.harvard.edu/abs/2014Natur.513...71T}
}

@ARTICLE{Hoffman_2017,
       author = {{Hoffman}, Yehuda and {Pomar{\`e}de}, Daniel and {Tully}, R. Brent and {Courtois}, H{\'e}l{\`e}ne M.},
        title = "{The dipole repeller}",
      journal = {Nature Astronomy},
         year = 2017,
        month = jan,
       volume = {1},
          eid = {0036},
        pages = {0036},
          doi = {10.1038/s41550-016-0036},
archivePrefix = {arXiv},
       eprint = {1702.02483},
 primaryClass = {astro-ph.CO},
       adsurl = {https://ui.adsabs.harvard.edu/abs/2017NatAs...1E..36H}
}

@ARTICLE{Aaronson_1982,
       author = {{Aaronson}, M. and {Huchra}, J. and {Mould}, J. and {Schechter}, P.~L. and {Tully}, R.~B.},
        title = "{The velocity field in the local supercluster.}",
      journal = {\apj},
         year = 1982,
        month = jul,
       volume = {258},
        pages = {64-76},
          doi = {10.1086/160053},
       adsurl = {https://ui.adsabs.harvard.edu/abs/1982ApJ...258...64A}
}

@ARTICLE{Bertschinger_1988,
       author = {{Bertschinger}, Edmund and {Juszkiewicz}, Roman},
        title = "{Searching for the Great Attractor}",
      journal = {\apjl},
         year = 1988,
        month = nov,
       volume = {334},
        pages = {L59},
          doi = {10.1086/185312},
       adsurl = {https://ui.adsabs.harvard.edu/abs/1988ApJ...334L..59B}
}

@ARTICLE{Dressler_1990,
       author = {{Dressler}, Alan and {Faber}, S.~M.},
        title = "{Confirmation of a Large-Scale, Large-Amplitude Flow in the Direction of the Great Attractor}",
      journal = {\apj},
         year = 1990,
        month = may,
       volume = {354},
        pages = {13},
          doi = {10.1086/168663},
       adsurl = {https://ui.adsabs.harvard.edu/abs/1990ApJ...354...13D}
}

@ARTICLE{Dressler_1990B,
       author = {{Dressler}, Alan and {Faber}, S.~M.},
        title = "{New Measurements of Distances to Spirals in the Great Attractor: Further Confirmation of the Large-Scale Flow}",
      journal = {\apjl},
         year = 1990,
        month = may,
       volume = {354},
        pages = {L45},
          doi = {10.1086/185719},
       adsurl = {https://ui.adsabs.harvard.edu/abs/1990ApJ...354L..45D}
}

@ARTICLE{Burstein_1990,
       author = {{Burstein}, David and {Faber}, S.~M. and {Dressler}, Alan},
        title = "{Evidence from the Motions of Galaxies for a Large-Scale, Large-Amplitude Flow toward the Great Attractor}",
      journal = {\apj},
         year = 1990,
        month = may,
       volume = {354},
        pages = {18},
          doi = {10.1086/168664},
       adsurl = {https://ui.adsabs.harvard.edu/abs/1990ApJ...354...18B}
}

@ARTICLE{Scaramella_1989,
       author = {{Scaramella}, R. and {Baiesi-Pillastrini}, G. and {Chincarini}, G. and {Vettolani}, G. and {Zamorani}, G.},
        title = "{A marked concentration of galaxy clusters: is this the origin of large-scale motions?}",
      journal = {\nat},
         year = 1989,
        month = apr,
       volume = {338},
       number = {6216},
        pages = {562-564},
          doi = {10.1038/338562a0},
       adsurl = {https://ui.adsabs.harvard.edu/abs/1989Natur.338..562S}
}

@ARTICLE{Raychaudhury_1989,
       author = {{Raychaudhury}, Somak},
        title = "{The distribution of galaxies in the direction of the 'Great Attractor'}",
      journal = {\nat},
         year = 1989,
        month = nov,
       volume = {342},
       number = {6247},
        pages = {251-255},
          doi = {10.1038/342251a0},
       adsurl = {https://ui.adsabs.harvard.edu/abs/1989Natur.342..251R}
}

@ARTICLE{Kraan-Korteweg_1996,
       author = {{Kraan-Korteweg}, R.~C. and {Woudt}, P.~A. and {Cayatte}, V. and {Fairall}, A.~P. and {Balkowski}, C. and {Henning}, P.~A.},
        title = "{A nearby massive galaxy cluster behind the Milky Way}",
      journal = {\nat},
         year = 1996,
        month = feb,
       volume = {379},
       number = {6565},
        pages = {519-521},
          doi = {10.1038/379519a0},
       adsurl = {https://ui.adsabs.harvard.edu/abs/1996Natur.379..519K}
}

@ARTICLE{Courtois_2013,
       author = {{Courtois}, H{\'e}l{\`e}ne M. and {Pomar{\`e}de}, Daniel and {Tully}, R. Brent and {Hoffman}, Yehuda and {Courtois}, Denis},
        title = "{Cosmography of the Local Universe}",
      journal = {\aj},
         year = 2013,
        month = sep,
       volume = {146},
       number = {3},
          eid = {69},
        pages = {69},
          doi = {10.1088/0004-6256/146/3/69},
archivePrefix = {arXiv},
       eprint = {1306.0091},
 primaryClass = {astro-ph.CO},
       adsurl = {https://ui.adsabs.harvard.edu/abs/2013AJ....146...69C}
}

@ARTICLE{CF3_cosmography,
       author = {{Tully}, R. Brent and {Pomar{\`e}de}, Daniel and {Graziani}, Romain and {Courtois}, H{\'e}l{\`e}ne M. and {Hoffman}, Yehuda and {Shaya}, Edward J.},
        title = "{Cosmicflows-3: Cosmography of the Local Void}",
      journal = {\apj},
         year = 2019,
        month = jul,
       volume = {880},
       number = {1},
          eid = {24},
        pages = {24},
          doi = {10.3847/1538-4357/ab2597},
archivePrefix = {arXiv},
       eprint = {1905.08329},
 primaryClass = {astro-ph.CO},
       adsurl = {https://ui.adsabs.harvard.edu/abs/2019ApJ...880...24T}
}

@ARTICLE{Hudson_1993_A,
       author = {{Hudson}, M.~J.},
        title = "{Optical galaxies within 8000 kms -1- I. The density field.}",
      journal = {\mnras},
         year = 1993,
        month = nov,
       volume = {265},
        pages = {43-71},
          doi = {10.1093/mnras/265.1.43},
       adsurl = {https://ui.adsabs.harvard.edu/abs/1993MNRAS.265...43H}
}

@ARTICLE{Hudson_1994_D,
       author = {{Hudson}, M.~J.},
        title = "{Optical galaxies within 8000 KM s-1 - IV. The peculiar velocity field.}",
      journal = {\mnras},
         year = 1994,
        month = jan,
       volume = {266},
        pages = {475-488},
          doi = {10.1093/mnras/266.2.475},
       adsurl = {https://ui.adsabs.harvard.edu/abs/1994MNRAS.266..475H}
}

@ARTICLE{Hudson_1993_B,
       author = {{Hudson}, M.~J.},
        title = "{Optical galaxies within 8000 KM s-1- II. The peculiar velocity of the Local Group.}",
      journal = {\mnras},
         year = 1993,
        month = nov,
       volume = {265},
        pages = {72-80},
          doi = {10.1093/mnras/265.1.72},
       adsurl = {https://ui.adsabs.harvard.edu/abs/1993MNRAS.265...72H}
}

@ARTICLE{Hudson_1994_C,
       author = {{Hudson}, M.~J.},
        title = "{Optical galaxies within 8000 km s$^{-1}$. III. Inhomogeneous Malmquist bias corrections and the Great Attractor.}",
      journal = {\mnras},
         year = 1994,
        month = jan,
       volume = {266},
        pages = {468-474},
          doi = {10.1093/mnras/266.2.468},
       adsurl = {https://ui.adsabs.harvard.edu/abs/1994MNRAS.266..468H}
}

@ARTICLE{KraanKorteweg_2017,
       author = {{Kraan-Korteweg}, Ren{\'e}e C. and {Cluver}, Michelle E. and {Bilicki}, Maciej and {Jarrett}, Thomas H. and {Colless}, Matthew and {Elagali}, Ahmed and {B{\"o}hringer}, Hans and {Chon}, Gayoung},
        title = "{Discovery of a supercluster in the Zone of Avoidance in Vela}",
      journal = {\mnras},
         year = 2017,
        month = mar,
       volume = {466},
       number = {1},
        pages = {L29-L33},
          doi = {10.1093/mnrasl/slw229},
archivePrefix = {arXiv},
       eprint = {1611.04615},
 primaryClass = {astro-ph.CO},
       adsurl = {https://ui.adsabs.harvard.edu/abs/2017MNRAS.466L..29K}
}

@ARTICLE{Courtois_2019,
       author = {{Courtois}, H{\'e}l{\`e}ne M. and {Kraan-Korteweg}, Ren{\'e}e C. and {Dupuy}, Alexandra and {Graziani}, Romain and {Libeskind}, Noam I.},
        title = "{A kinematic confirmation of the hidden Vela supercluster}",
      journal = {\mnras},
         year = 2019,
        month = nov,
       volume = {490},
       number = {1},
        pages = {L57-L61},
          doi = {10.1093/mnrasl/slz146},
archivePrefix = {arXiv},
       eprint = {1909.08875},
 primaryClass = {astro-ph.CO},
       adsurl = {https://ui.adsabs.harvard.edu/abs/2019MNRAS.490L..57C}
}

@ARTICLE{Hatamkhani_2023,
       author = {{Hatamkhani}, N. and {Kraan-Korteweg}, R.~C. and {Blyth}, S.~L. and {Said}, K. and {Elagali}, A.},
        title = "{Galaxy clusters in the Vela Supercluster - I. Deep NIR catalogues}",
      journal = {\mnras},
         year = 2023,
        month = jun,
       volume = {522},
       number = {2},
        pages = {2223-2240},
          doi = {10.1093/mnras/stad1134},
archivePrefix = {arXiv},
       eprint = {2304.07208},
 primaryClass = {astro-ph.GA},
       adsurl = {https://ui.adsabs.harvard.edu/abs/2023MNRAS.522.2223H}
}

@ARTICLE{Rajohnson_2024,
       author = {{Rajohnson}, Sambatriniaina H.~A. and {Kraan-Korteweg}, Ren{\'e}e C. and {Frank}, Bradley S. and {Chen}, Hao and {Staveley-Smith}, Lister and {Serra}, Paolo and {Steyn}, Nadia and {Kurapati}, Sushma and {Pisano}, D.~J. and {Goedhart}, Sharmila},
        title = "{Revealing hidden structures in the Zone of Avoidance - a blind MeerKAT H I Survey of the Vela Supercluster}",
      journal = {\mnras},
         year = 2024,
        month = dec,
       volume = {535},
       number = {4},
        pages = {3429-3450},
          doi = {10.1093/mnras/stae2552},
archivePrefix = {arXiv},
       eprint = {2411.07084},
 primaryClass = {astro-ph.GA},
       adsurl = {https://ui.adsabs.harvard.edu/abs/2024MNRAS.535.3429R}
}

@ARTICLE{Hoffman_2024,
       author = {{Hoffman}, Yehuda and {Valade}, Aurelien and {Libeskind}, Noam I. and {Sorce}, Jenny G. and {Tully}, R. Brent and {Pfeifer}, Simon and {Gottl{\"o}ber}, Stefan and {Pomar{\`e}de}, Daniel},
        title = "{The large-scale velocity field from the Cosmicflows-4 data}",
      journal = {\mnras},
         year = 2024,
        month = jan,
       volume = {527},
       number = {2},
        pages = {3788-3805},
          doi = {10.1093/mnras/stad3433},
archivePrefix = {arXiv},
       eprint = {2311.01340},
 primaryClass = {astro-ph.CO},
       adsurl = {https://ui.adsabs.harvard.edu/abs/2024MNRAS.527.3788H}
}

@ARTICLE{Courtois_2025,
       author = {{Courtois}, H.~M. and {Mould}, J. and {Hollinger}, A.~M. and {Dupuy}, A. and {Zhang}, C.~P.},
        title = "{In search of the Local Universe dynamical homogeneity scale with CF4++ peculiar velocities}",
      journal = {\aap},
         year = 2025,
        month = sep,
       volume = {701},
          eid = {A187},
        pages = {A187},
          doi = {10.1051/0004-6361/202553677},
archivePrefix = {arXiv},
       eprint = {2502.01308},
 primaryClass = {astro-ph.CO},
       adsurl = {https://ui.adsabs.harvard.edu/abs/2025A&A...701A.187C}
}

@ARTICLE{Courtois_2023,
       author = {{Courtois}, H.~M. and {Dupuy}, A. and {Guinet}, D. and {Baulieu}, G. and {Ruppin}, F. and {Brenas}, P.},
        title = "{Gravity in the local Universe: Density and velocity fields using CosmicFlows-4}",
      journal = {\aap},
         year = 2023,
        month = feb,
       volume = {670},
          eid = {L15},
        pages = {L15},
          doi = {10.1051/0004-6361/202245331},
archivePrefix = {arXiv},
       eprint = {2211.16390},
 primaryClass = {astro-ph.CO},
       adsurl = {https://ui.adsabs.harvard.edu/abs/2023A&A...670L..15C}
}

@ARTICLE{Tully_2023,
       author = {{Tully}, R. Brent and {Kourkchi}, Ehsan and {Courtois}, H{\'e}l{\`e}ne M. and {Anand}, Gagandeep S. and {Blakeslee}, John P. and {Brout}, Dillon and {Jaeger}, Thomas de and {Dupuy}, Alexandra and {Guinet}, Daniel and {Howlett}, Cullan and {Jensen}, Joseph B. and {Pomar{\`e}de}, Daniel and {Rizzi}, Luca and {Rubin}, David and {Said}, Khaled and {Scolnic}, Daniel and {Stahl}, Benjamin E.},
        title = "{Cosmicflows-4}",
      journal = {\apj},
         year = 2023,
        month = feb,
       volume = {944},
       number = {1},
          eid = {94},
        pages = {94},
          doi = {10.3847/1538-4357/ac94d8},
archivePrefix = {arXiv},
       eprint = {2209.11238},
 primaryClass = {astro-ph.CO},
       adsurl = {https://ui.adsabs.harvard.edu/abs/2023ApJ...944...94T}
}

@ARTICLE{Wempe_2024,
       author = {{Wempe}, Ewoud and {Lavaux}, Guilhem and {White}, Simon D.~M. and {Helmi}, Amina and {Jasche}, Jens and {Stopyra}, Stephen},
        title = "{Constrained cosmological simulations of the Local Group using Bayesian hierarchical field-level inference}",
      journal = {\aap},
         year = 2024,
        month = nov,
       volume = {691},
          eid = {A348},
        pages = {A348},
          doi = {10.1051/0004-6361/202450975},
archivePrefix = {arXiv},
       eprint = {2406.02228},
 primaryClass = {astro-ph.GA},
       adsurl = {https://ui.adsabs.harvard.edu/abs/2024A&A...691A.348W}
}

@ARTICLE{Wempe_2025,
       author = {{Wempe}, Ewoud and {Helmi}, Amina and {White}, Simon D.~M. and {Jasche}, Jens and {Lavaux}, Guilhem},
        title = "{The effect of environment on the mass assembly history of the Milky Way and M31}",
      journal = {\aap},
         year = 2025,
        month = sep,
       volume = {701},
          eid = {A178},
        pages = {A178},
          doi = {10.1051/0004-6361/202553744},
archivePrefix = {arXiv},
       eprint = {2501.08089},
 primaryClass = {astro-ph.GA},
       adsurl = {https://ui.adsabs.harvard.edu/abs/2025A&A...701A.178W}
}

@ARTICLE{Kourkchi_2020,
       author = {{Kourkchi}, Ehsan and {Tully}, R. Brent and {Anand}, Gagandeep S. and {Courtois}, H{\'e}l{\`e}ne M. and {Dupuy}, Alexandra and {Neill}, James D. and {Rizzi}, Luca and {Seibert}, Mark},
        title = "{Cosmicflows-4: The Calibration of Optical and Infrared Tully-Fisher Relations}",
      journal = {\apj},
         year = 2020,
        month = jun,
       volume = {896},
       number = {1},
          eid = {3},
        pages = {3},
          doi = {10.3847/1538-4357/ab901c},
archivePrefix = {arXiv},
       eprint = {2004.14499},
 primaryClass = {astro-ph.GA},
       adsurl = {https://ui.adsabs.harvard.edu/abs/2020ApJ...896....3K}
}

@ARTICLE{Quijote,
       author = {{Villaescusa-Navarro}, Francisco and {Hahn}, ChangHoon and {Massara}, Elena and {Banerjee}, Arka and {Delgado}, Ana Maria and {Ramanah}, Doogesh Kodi and {Charnock}, Tom and {Giusarma}, Elena and {Li}, Yin and {Allys}, Erwan and {Brochard}, Antoine and {Uhlemann}, Cora and {Chiang}, Chi-Ting and {He}, Siyu and {Pisani}, Alice and {Obuljen}, Andrej and {Feng}, Yu and {Castorina}, Emanuele and {Contardo}, Gabriella and {Kreisch}, Christina D. and {Nicola}, Andrina and {Alsing}, Justin and {Scoccimarro}, Roman and {Verde}, Licia and {Viel}, Matteo and {Ho}, Shirley and {Mallat}, Stephane and {Wandelt}, Benjamin and {Spergel}, David N.},
        title = "{The Quijote Simulations}",
      journal = {\apjs},
         year = 2020,
        month = sep,
       volume = {250},
       number = {1},
          eid = {2},
        pages = {2},
          doi = {10.3847/1538-4365/ab9d82},
archivePrefix = {arXiv},
       eprint = {1909.05273},
 primaryClass = {astro-ph.CO},
       adsurl = {https://ui.adsabs.harvard.edu/abs/2020ApJS..250....2V}
}

@ARTICLE{Springel_2005,
       author = {{Springel}, Volker},
        title = "{The cosmological simulation code GADGET-2}",
      journal = {\mnras},
         year = 2005,
        month = dec,
       volume = {364},
       number = {4},
        pages = {1105-1134},
          doi = {10.1111/j.1365-2966.2005.09655.x},
archivePrefix = {arXiv},
       eprint = {astro-ph/0505010},
 primaryClass = {astro-ph},
       adsurl = {https://ui.adsabs.harvard.edu/abs/2005MNRAS.364.1105S}
}

@ARTICLE{Bouchet_1995,
       author = {{Bouchet}, F.~R. and {Colombi}, S. and {Hivon}, E. and {Juszkiewicz}, R.},
        title = "{Perturbative Lagrangian approach to gravitational instability.}",
      journal = {\aap},
         year = 1995,
        month = apr,
       volume = {296},
        pages = {575},
          doi = {10.48550/arXiv.astro-ph/9406013},
archivePrefix = {arXiv},
       eprint = {astro-ph/9406013},
 primaryClass = {astro-ph},
       adsurl = {https://ui.adsabs.harvard.edu/abs/1995A&A...296..575B}
}

@ARTICLE{Wang_1998,
       author = {{Wang}, Limin and {Steinhardt}, Paul J.},
        title = "{Cluster Abundance Constraints for Cosmological Models with a Time-varying, Spatially Inhomogeneous Energy Component with Negative Pressure}",
      journal = {\apj},
         year = 1998,
        month = dec,
       volume = {508},
       number = {2},
        pages = {483-490},
          doi = {10.1086/306436},
archivePrefix = {arXiv},
       eprint = {astro-ph/9804015},
 primaryClass = {astro-ph},
       adsurl = {https://ui.adsabs.harvard.edu/abs/1998ApJ...508..483W}
}

@BOOK{Peebles_1980,
       author = {{Peebles}, P.~J.~E.},
        title = "{The large-scale structure of the universe}",
         year = 1980,
       adsurl = {https://ui.adsabs.harvard.edu/abs/1980lssu.book.....P}
}

@ARTICLE{Lavaux_2010,
       author = {{Lavaux}, Guilhem and {Tully}, R. Brent and {Mohayaee}, Roya and {Colombi}, St{\'e}phane},
        title = "{Cosmic Flow From Two Micron All-Sky Redshift Survey: the Origin of Cosmic Microwave Background Dipole and Implications for {\ensuremath{\Lambda}}CDM Cosmology}",
      journal = {\apj},
         year = 2010,
        month = jan,
       volume = {709},
       number = {1},
        pages = {483-498},
          doi = {10.1088/0004-637X/709/1/483},
archivePrefix = {arXiv},
       eprint = {0810.3658},
 primaryClass = {astro-ph},
       adsurl = {https://ui.adsabs.harvard.edu/abs/2010ApJ...709..483L}
}

@ARTICLE{Carrick_2015,
       author = {{Carrick}, Jonathan and {Turnbull}, Stephen J. and {Lavaux}, Guilhem and {Hudson}, Michael J.},
        title = "{Cosmological parameters from the comparison of peculiar velocities with predictions from the 2M++ density field}",
      journal = {\mnras},
         year = 2015,
        month = jun,
       volume = {450},
       number = {1},
        pages = {317-332},
          doi = {10.1093/mnras/stv547},
archivePrefix = {arXiv},
       eprint = {1504.04627},
 primaryClass = {astro-ph.CO},
       adsurl = {https://ui.adsabs.harvard.edu/abs/2015MNRAS.450..317C}
}

@ARTICLE{Benisty_2022_LGmass,
       author = {{Benisty}, David and {Vasiliev}, Eugene and {Evans}, N. Wyn and {Davis}, Anne-Christine and {Hartl}, Odelia V. and {Strigari}, Louis E.},
        title = "{The Local Group Mass in the Light of Gaia}",
      journal = {\apjl},
         year = 2022,
        month = mar,
       volume = {928},
       number = {1},
          eid = {L5},
        pages = {L5},
          doi = {10.3847/2041-8213/ac5c42},
archivePrefix = {arXiv},
       eprint = {2202.00033},
 primaryClass = {astro-ph.GA},
       adsurl = {https://ui.adsabs.harvard.edu/abs/2022ApJ...928L...5B}
}

@ARTICLE{Pike_2005,
       author = {{Pike}, R.~W. and {Hudson}, Michael J.},
        title = "{Cosmological Parameters from the Comparison of the 2MASS Gravity Field with Peculiar Velocity Surveys}",
      journal = {\apj},
         year = 2005,
        month = dec,
       volume = {635},
       number = {1},
        pages = {11-21},
          doi = {10.1086/497359},
archivePrefix = {arXiv},
       eprint = {astro-ph/0511012},
 primaryClass = {astro-ph},
       adsurl = {https://ui.adsabs.harvard.edu/abs/2005ApJ...635...11P}
}

@ARTICLE{Peacock_1992,
       author = {{Peacock}, J.~A.},
        title = "{Errors on the measurement of omega via cosmological dipoles.}",
      journal = {\mnras},
         year = 1992,
        month = oct,
       volume = {258},
        pages = {581-586},
          doi = {10.1093/mnras/258.3.581},
       adsurl = {https://ui.adsabs.harvard.edu/abs/1992MNRAS.258..581P}
}

@ARTICLE{Lilje_1986,
       author = {{Lilje}, P.~B. and {Yahil}, A. and {Jones}, B.~J.~T.},
        title = "{The Tidal Velocity Field in the Local Supercluster}",
      journal = {\apj},
         year = 1986,
        month = aug,
       volume = {307},
        pages = {91},
          doi = {10.1086/164395},
       adsurl = {https://ui.adsabs.harvard.edu/abs/1986ApJ...307...91L}
}

\appendix

\section{Further visualisation of the cGA region}\label{sec:appendix_ga_figs}

\begin{figure*}
    \centering
    \includegraphics[width=0.9\textwidth]{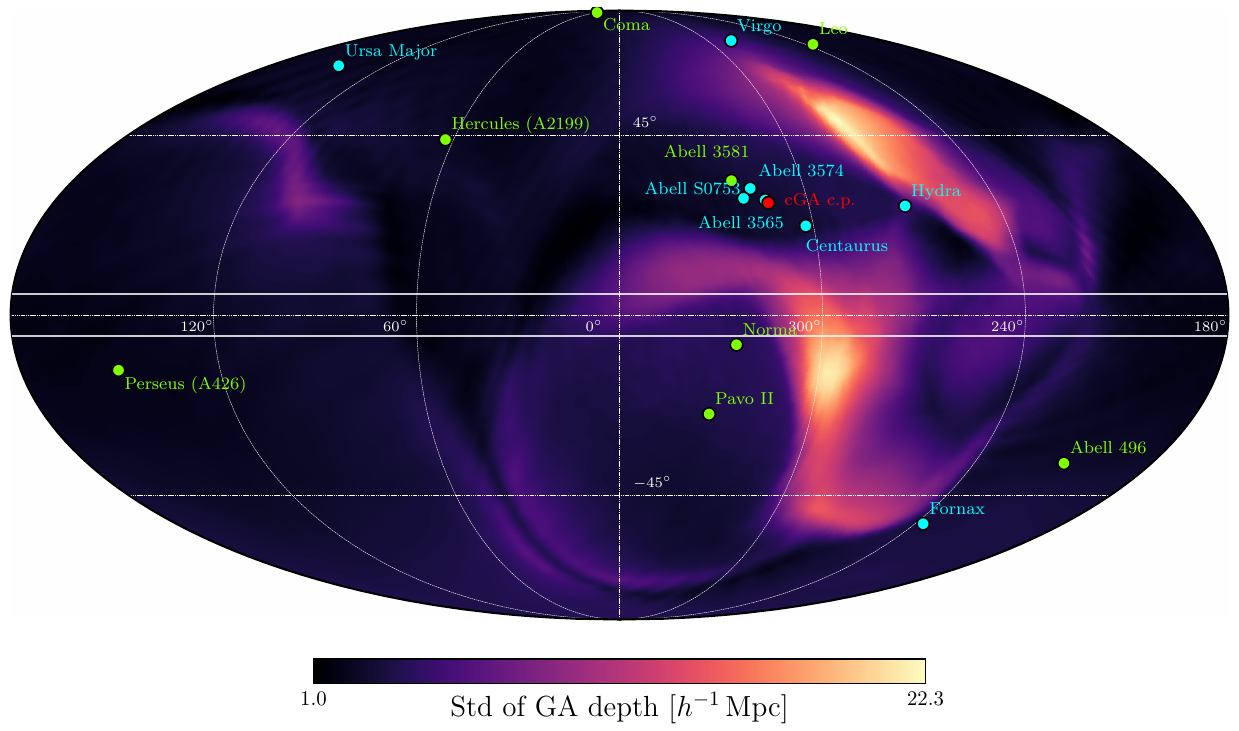}
    \caption{
        Standard deviation of the \ac{cGA} basin depth across 80 \Manti realisations at $\sigma_{\rm smooth}=3~\Mpch$, complementing the mean depth shown in~\cref{fig:ga_depth_sky}.
        Variance is smallest toward the \ac{cGA} direction and largest along the basin boundary, highlighting where the attractor's extent is most and least certain; the \ac{ZoA} (white lines at $b = \pm 5\,\deg$) is retained for reference.
        Dashed white lines indicate meridians every $60\,\deg$ and parallels at $\pm 45\,\deg$.
        }
    \label{fig:ga_depth_sky_std}
\end{figure*}

\begin{figure*}
    \centering
    \includegraphics[width=0.9\textwidth]{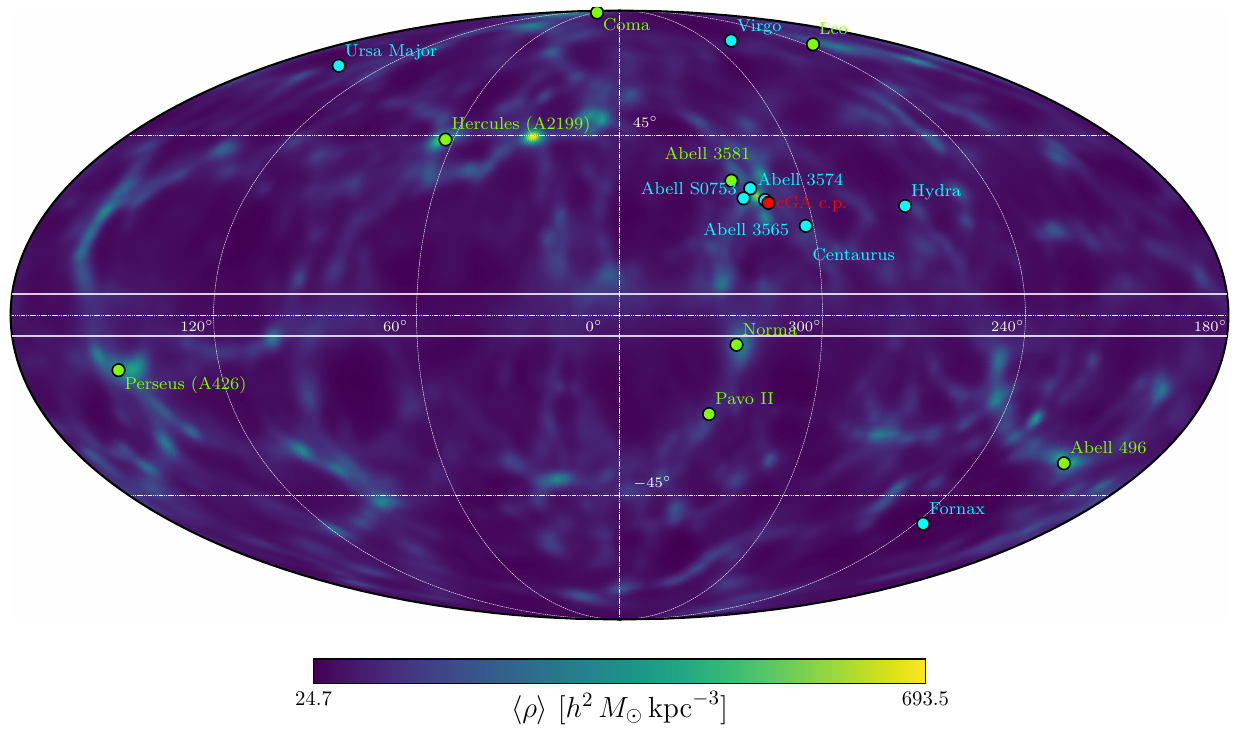}
    \caption{
        Sky-projected mean density field at $\sigma_{\rm smooth}=3~\Mpch$, computed by averaging the 80 \Manti realisations and, for each sky pixel, evaluating $\langle \rho \rangle_{\rm sky} = \bigl[\int_{0}^{80~\Mpch} r^{2} \rho(r,\,\ell,\,b)\,\mathrm{d}r\bigr]/\bigl[\int_{0}^{80~\Mpch} r^{2}\,\mathrm{d}r\bigr]$.
        The map highlights the projected overdensities associated with the \ac{cGA} region and surrounding structures in Galactic coordinates; density within the \ac{ZoA} (white lines at $b = \pm 5\,\deg$) appears blurrier, though coherent structure persists across it.
        Dashed white lines indicate meridians every $60\,\deg$ and parallels at $\pm 45\,\deg$.
        }
    \label{fig:sky_density}
\end{figure*}

Here we briefly present additional figures visualising the \ac{cGA} region identified in \Manti, to supplement the main text.
Firstly, in \cref{fig:ga_depth_sky_std} we show the standard deviation of the \ac{cGA} basin depth across the 80 \Manti realisations, complementing the mean depth shown in \cref{fig:ga_depth_sky}.
Secondly, \cref{fig:sky_density} presents the sky-projected mean density field, obtained by averaging the 80 \Manti realisations and, for each pixel, computing
\begin{equation}
    \langle \rho \rangle_{\rm sky} = \frac{\int_{0}^{80~\Mpch} r^{2} \rho(r,\,\ell,\,b)\,\mathrm{d}r}{\int_{0}^{80~\Mpch} r^{2}\,\mathrm{d}r},
    \vspace{1em}
\end{equation}
which integrates $r^{2}\rho$ along the line of sight to $80~\Mpch$ (approximately the maximal \ac{cGA} extent from the observer) and normalises by the corresponding volume factor.

\end{document}